\pdfoutput=1

\documentclass[11pt]{article}

\usepackage[preprint]{acl}

\usepackage{times}
\usepackage{latexsym}
\usepackage{booktabs}
\usepackage{multirow}
\usepackage{graphicx}
\usepackage{subcaption} 
\usepackage{array}
\usepackage{algorithm}
\usepackage{algpseudocode}
\usepackage{amsmath}
\usepackage{graphicx}
\usepackage{amssymb}
\usepackage{hyperref}
\usepackage[T1]{fontenc}

\usepackage[utf8]{inputenc}

\usepackage{microtype}

\usepackage{inconsolata}

\usepackage{graphicx}

\title{Learning Explainable Stock Predictions with Tweets Using Mixture of Experts}

\author{
Wenyan Xu$^{1}$\thanks{\quad Equal contribution} \quad Dawei Xiang$^{2}$\footnotemark[1] \quad Rundong Wang$^{3}$ \\ 
{\bf Yonghong Hu}$^{1}$ \quad {\bf Liang Zhang}$^{4}$ \quad {\bf Jiayu Chen}$^{5}$ \quad {\bf Zhonghua Lu}$^{6}$ \\
\\
\textsuperscript{1}Central University of Finance and Economics \quad 
\textsuperscript{2}University of Connecticut \\
\textsuperscript{3}Nanyang Technological University \quad
\textsuperscript{4}The Hong Kong University of Science and Technology (Guangzhou) \\
\textsuperscript{5}Hong Kong University \quad
\textsuperscript{6}University of Chinese Academy of Sciences
}

\begin{document}
\maketitle
\begin{abstract}
Stock price movements are influenced by many factors, and alongside historical price data, textual information is a key source. Public news and social media offer valuable insights into market sentiment and emerging events. These sources are fast-paced, diverse, and significantly impact future stock trends. Recently, LLMs have enhanced financial analysis, but prompt-based methods still have limitations, such as input length restrictions and difficulties in predicting sequences of varying lengths. Additionally, most models rely on dense computational layers, which are resource-intensive. To address these challenges, we propose the FTS-Text-MoE model, which combines numerical data with key summaries from news and tweets using point embeddings, boosting prediction accuracy through the integration of factual textual data. The model uses a Mixture of Experts (MoE) Transformer decoder to process both data types. By activating only a subset of model parameters, it reduces computational costs. Furthermore, the model features multi-resolution prediction heads, enabling flexible forecasting of financial time series at different scales. Experimental results show that FTS-Text-MoE outperforms baseline methods in terms of investment returns and Sharpe ratio, demonstrating its superior accuracy and ability to predict future market trends. 
\end{abstract}

\section{Introduction}

Stock price prediction is inherently a time series problem, and time series regression models have long been central to financial valuation. Traditional methods such as linear regression\cite{montgomery2021introduction}, ARIMA\cite{ariyo2014stock}, and GARCH\cite{francq2019garch} typically assume long-term market stability and rely on predefined assumptions. However, these methods are limited in their ability to capture complex market dependencies and adapt to sudden market events (Malkiel, 1999). Recent studies have highlighted the advantages of machine learning techniques, which excel at modeling long-term dependencies and detecting market volatility, overcoming the reliance on prior assumptions typical of traditional methods\cite{kumar2024dynamic}.

Unlike traditional models that rely solely on numerical data, deep learning models\cite{koa2023diffusion} can integrate multi-dimensional, heterogeneous information to improve prediction accuracy. Harry Markowitz's Modern Portfolio Theory\cite{konstantinov2020network} emphasizes the importance of market correlations, while research by \cite{hsu2021news} demonstrates a positive correlation between emotions in news, blogs, and social media and stock market trends\cite{hsu2021news}. However, relying solely on sentiment scores may overlook key details in text content, failing to fully tap into its potential. Therefore, integrating richer and more comprehensive text data into stock prediction models has become increasingly important.


Recent cross-modal time series prediction methods based on large language models (LLMs) have demonstrated outstanding performance. In the latest portfolio optimization applications, SocioDojo\cite{cheng2024sociodojo} and SEP\cite{koa2024learning} use knowledge-aligned text prompts to help LLMs make more accurate decisions. However, these methods are limited by the input context length, processing only a limited amount of text, and have a restricted prediction horizon. Furthermore, they typically perform binary classification to predict stock price trends rather than accurately forecasting the next value in the time series. Additionally, the computational cost of invoking LLMs is significant. Unlike the aforementioned prompt-based methods, we sort news and tweet data by time and extract 1-2 key factual summaries at each time point, aligning them with numerical time series via point embeddings. Additionally, we predict the actual stock price value, rather than just the trend.

Most time series prediction models rely on dense layers, requiring the computation of all parameters for each input token. Although these methods are computationally accurate, they consume substantial resources. Sparse techniques, such as the Mixture of Experts (MoE), improve computational efficiency and model performance under fixed inference constraints. Based on this, we propose the Financial Time Series-Text-MoE (FTS-Text-MoE) model, a decoder-only transformer architecture specifically designed for stock price prediction. The model uses a sparse transformer decoder, selectively activating parameters to significantly reduce computational costs. Additionally, we design multi-resolution prediction heads to enable more flexible predictions for time series of varying lengths. 

Key contributions of this study include: 
\begin{itemize} \setlength{\itemsep}{0pt}
\item We resolved technical issues in crawling public news data and updated the Nasdaq news dataset in the FNSPID project \footnote{\scriptsize\url{https://github.com/Zdong104/FNSPID_Financial_News_Dataset}} to January 19, 2025 (the original dataset was until 2023), detailed in Appendix \ref{News Data}. 
\item We introduced the FTS-Text-MoE model 
, specifically designed for stock price prediction, which effectively aligns time series data with text data, improving prediction accuracy.
\item We demonstrated through experiments that FTS-Text-MoE outperforms traditional benchmark methods in accurately predicting financial time series trends, with significant improvements in investment returns and Sharpe ratio. 
\end{itemize} 

\begin{figure*}[t]
  \centering
\includegraphics[width=1\textwidth]{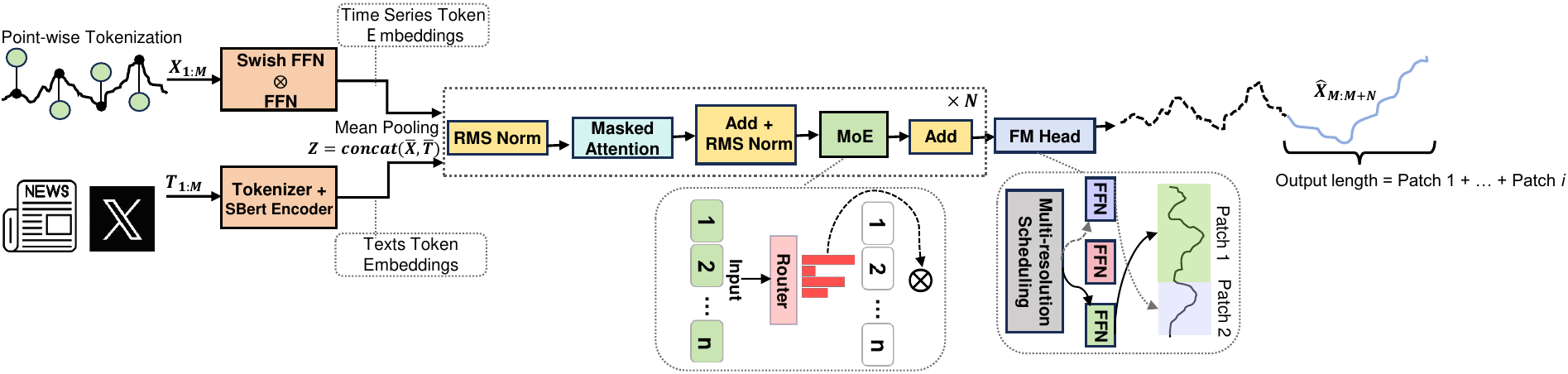}
    \vspace{-2mm}
  \caption{This figure presents the architecture of FTS-Text-MoE, a model designed to integrate textual and time series data for forecasting. }
  \vspace{-5mm}
  \label{fig:framework}
\end{figure*}

\section{Related Works}
Early studies on stock prediction using text analysis employed support vector machines (SVM) to extract basic features from text, such as bag-of-words, noun phrases, and named entities \cite{schumaker2009textual}. These shallow features were later replaced by structured tuples (e.g., subject, action, object) extracted by deep neural networks \cite{ding2015deep, deng2019knowledge}. Additionally, \cite{xu2018stock} extracted implicit text vectors and used variational autoencoders (VAE) to model stock price prediction as a binary classification problem. As understanding of text complexity grew, \cite{xu2018stock, yang2020html, deng2019knowledge} began extracting richer implicit information from pre-trained text embeddings. Recent studies, such as \cite{wang2025chattime, liu2025timecma}, used text as prompts to further enhance the accuracy of LLMs in time-series prediction. Unlike these methods, our research couples text sentences at each time point with corresponding time series data, predicting the next time series value, providing clearer analysis and exploring the specific impact of text on stock price fluctuations.

\textbf{Sparse Deep Learning for Time Series.} Deep learning models are typically large-scale and parameter-intensive \cite{hoefler2021sparsity}, requiring significant memory and computational resources. Sparse networks, such as MoE \cite{jacobs1991adaptive}, use dynamic routing to specialized experts, improving efficiency while maintaining or exceeding generalization performance \cite{fedus2022switch}. Traditional time series models, like DLinear \cite{zeng2023transformers} and SparseTSF \cite{lin2024sparsetsf}, are usually smaller and have less focus on sparse methods. After large-scale pretraining, studies such as MoLE \cite{ni2024mixture} and IME \cite{ismailinterpretable} explore sparsity but are not fully sparse models, as they route inputs to all attention heads before aggregation. The recent study Time-MoE \cite{shi2024time} uses sparse foundational models for general time-series forecasting but does not consider the impact of real-world textual information on numerical features. We further investigate whether complex textual information can enhance the interpretability of numerical predictions.

\section{Methodology}

We propose FTS-Text-MoE (Figure \ref{fig:framework}), a decoder-only Transformer architecture utilizing an MoE framework. The model comprises three components: (1) input token embedding, (2) MoE Transformer module, and (3) multi-resolution forecasting.

\paragraph{Problem Formulation.} Given historical numerical observations $\mathbf{X}_{1:M} = \left(x_1, x_2, \ldots, x_M \right) \in \mathbb{R}^{M}$ and corresponding textual data $\mathbf{T}_{1:M} = \left(t_1, t_2, \ldots, t_M \right) \in \mathbb{R}^{M}$ over the past $M$ days, our goal is to predict the next $N$ days: $\hat{\mathbf{X}}_{M+1:M+N} = f_{\theta} \left( \mathbf{X}_{1:M}, \mathbf{T}_{1:M}\right) \in \mathbb{R}^{N}$ where $f_{\theta}$ represents the proposed model. FTS-Text-MoE supports flexible context lengths ($M$) and forecasting horizons ($N$) during inference, unlike traditional fixed-horizon models. Following the channel independence principle from \cite{nietime}, multivariate series are decomposed into univariate series, enabling better generalization across tasks.

\subsection{Input Token Embedding}

\begin{figure*}[t]
  \centering
\includegraphics[width=1\linewidth]{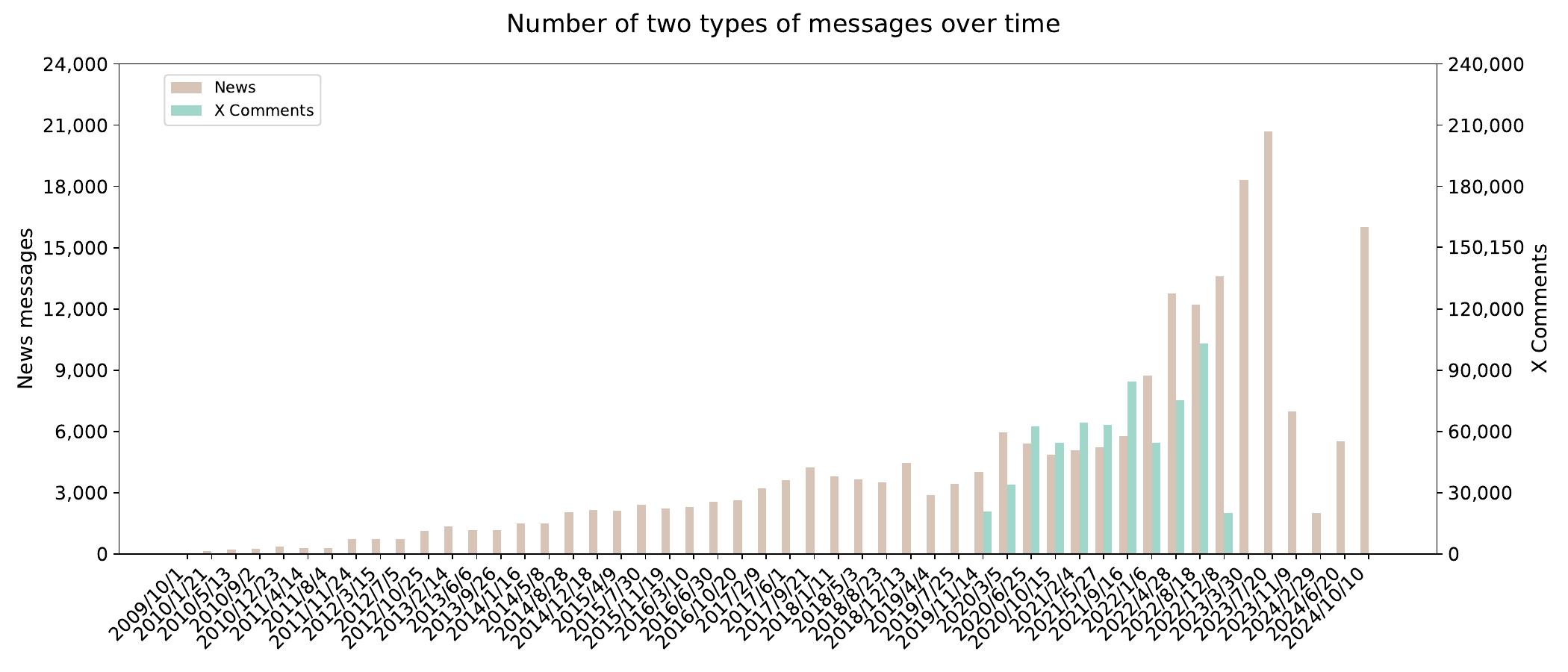}
  \vspace{-3mm}
  \caption{The x-axis represents dates, with each bar indicating the distribution of two text sources over the following 112 days.}
  \label{fig:tweet_news_nums}
\end{figure*}

\paragraph{Text Token Embedding} We present an efficient pipeline for converting raw text into high-quality embeddings. On a daily basis, we select the top $k$ news articles and X Comments (Figure \ref{fig:tweet_news_nums}), embedding them using a pretrained DistilBART-12-6-cnn model\footnote{\scriptsize\url{https://huggingface.co/sshleifer/distilbart-cnn-12-6}}. The resulting summaries are tokenized using the MiniLM L3 v2 (para) tokenizer\footnote{\scriptsize\url{https://huggingface.co/sentence-transformers/paraphrase-MiniLM-L3-v2}} and subsequently transformed into semantic representations using the SBertEncoder.

\paragraph{Time Token Embedding} Each time point in the financial time series is embedded using SwiGLU \cite{shazeer2020glu}:
\begin{equation}
\mathbf{h}_t^0 = \operatorname{SwiGLU}(x_t) = \operatorname{Swish}(W x_t) \otimes (V x_t),
\end{equation}
where  $W \in R^{D \times 1}$ and $V \in R^{D \times 1}$ and  are learnable parameters, and  denotes the hidden dimension.

Textual and numerical embeddings undergo mean pooling for aggregation before entering the subsequent MoE Transformer modules.

\begin{table*}[t]
    \centering
    \resizebox{\textwidth}{!}{ 
    \begin{tabular}{lcccccccccccccc}
        \toprule
        \multirow{2}{*}{Sector} & \multicolumn{2}{c}{Chronos} & \multicolumn{2}{c}{Moirai} & \multicolumn{2}{c}{$\text{Ours}_{\text{ts}}$} & \multicolumn{2}{c}{$\text{Ours}_{\text{ts\&news}}$} & \multicolumn{2}{c}{$\text{Ours}_{\text{ts\&x comments}}$}\\
        \cmidrule(lr){2-3} \cmidrule(lr){4-5} \cmidrule(lr){6-7} \cmidrule(lr){8-9}\cmidrule(lr){10-11}
        & MSE (\(\downarrow\)) & MAE (\(\downarrow\)) & MSE (\(\downarrow\)) & MAE (\(\downarrow\)) & MSE (\(\downarrow\)) & MAE (\(\downarrow\)) & MSE (\(\downarrow\)) & MAE (\(\downarrow\)) & MSE (\(\downarrow\)) & MAE (\(\downarrow\)) \\
        \midrule
        Basic Materials & 0.6458 & 0.8112 & 0.8725 & 0.8399 & 0.5540 & 0.6591 & \textbf{0.1972} & \textbf{0.3849} & 0.3866 & 0.5505 \\
        Financial Services & 0.7247 & 0.7549 & 0.7466 & 0.7644 & 0.5498 & 0.6429 & \textbf{0.2060} & \textbf{0.3913} & 0.2077 & 0.3914 \\
        Consumer Defensive & 0.1946 & 0.4503 & \textbf{0.1533} & \textbf{0.3558} & 0.5717 & 0.6721 & 0.5717 & 0.6721 & 0.5700 & 0.6710 \\
        Utilities & 0.7518 & 0.9498 & 0.2517 & 0.4080 & 0.2797 & 0.4188 & 0.2797 & 0.4188 & \textbf{0.0527} & \textbf{0.1698} \\
        Energy & 0.2094 & 0.3878 & 0.3204 & 0.4979 & 0.4196 & 0.5835 & 0.4196 & 0.5835 & \textbf{0.1392} & \textbf{0.3615} \\
        Technology & 0.8681 & 0.8893 & 0.5858 & 0.5920 & 0.7480 & 0.7412 & 0.8688 & 0.8279 & \textbf{0.1538} & \textbf{0.3643} \\
        Consumer Cyclical & 0.3628 & 0.5345 & 0.8673 & 0.9623 & 0.3116 & 0.5074 & 0.1632 & 0.3486 & \textbf{0.1049} & \textbf{0.2252} \\
        Real Estate & \textbf{0.2282} & \textbf{0.4495} & 0.3687 & 0.4488 & 0.3567 & 0.5419 & 0.3892 & 0.5461 & 0.3550 & 0.5420 \\
        Healthcare & 0.3521 & 0.4199 & 0.5786 & 0.6337 & 0.2248 & 0.4226 & \textbf{0.0705} & \textbf{0.2162} & 0.1299 & 0.3057 \\
        Communication Services & 0.2029 & 0.3351 & 0.7599 & 0.8566 & 0.3388 & 0.5166 & \textbf{0.0511} & \textbf{0.1817} & 0.0697 & 0.2225 \\
        Industrials & 0.4708 & 0.5574 & 0.2126 & 0.5158 & 0.1226 & 0.2743 & 0.5829 & 0.6761 & \textbf{0.1224} & \textbf{0.2741} \\
        \bottomrule
    \end{tabular}
    }
    \caption{Analysis of FTS-Text-MoE Performance with Mixed Text Inputs vs. Baselines}
    \vspace{-5mm}
    \label{tab:different-method-comparison}
\end{table*}

\subsection{FTS-Text-MoE Transformer}
Inspired by \cite{shi2024time}, we utilize a decoder-only Transformer \cite{vaswani2017attention, chen2024deep}, incorporating advancements from LLMs \cite{touvron2023llama}. To enhance training stability, we apply RMSNorm \cite{zhang2019root} at each sub-layer input. Rotary Positional Embeddings \cite{su2024roformer} improve sequence flexibility and extrapolation. 
Following \cite{chowdhery2023palm}, we eliminate most biases, retaining them only in the QKV layers of self-attention for better extrapolation. Formally, the architecture is defined as follows:
\begin{equation}
\resizebox{0.75\hsize}{!}{$
\begin{aligned}
\mathbf{a}_t^l &= \operatorname{MA}\left(\operatorname{RMSNorm}(\mathbf{x}_t^{l-1})\right) + \mathbf{x}_t^{l-1}, \\
\bar{\mathbf{a}}_t^l &= \operatorname{RMSNorm}(\mathbf{a}_t^l), \\
\mathbf{x}_t^l &= \operatorname{Mixture}(\bar{\mathbf{a}}_t^l) + \mathbf{a}_t^l.
\end{aligned}
$}
\label{equ:layer_operations}
\end{equation}
To introduce sparsity, we replace the standard FFN with a MoE layer, sharing a pool of sparsely activated experts. Each Mixture layer comprises multiple expert networks akin to standard FFNs, where each timestep routes inputs to either a single expert \cite{fedus2022switch} or multiple experts \cite{lepikhingshard}. One shared expert captures common knowledge across contexts. The Mixture function is defined as follows:

\resizebox{0.8\hsize}{!}{ \begin{minipage}{\linewidth} \begin{gather} \operatorname{Mixture}(\bar{\mathbf{a}}_t^l) = g_{N+1,t} \operatorname{FFN}_{N+1}(\bar{\mathbf{a}}_t^l) + \sum_{i=1}^{N} g_{i,t} \operatorname{FFN}_{i}(\bar{\mathbf{a}}_t^l) \label{equ:mixture-1}\\
\vspace{-2mm}
g_{i,t} = \begin{cases} s_{i,t}, & \text{if } s_{i,t} \in \operatorname{TopK}\left(\{s_{j,t}\mid 1 \leq j \leq N\}, K\right), \\[3pt] 0, & \text{otherwise}. \end{cases} \label{equ:mixture-3}\\ \vspace{-5mm} 
g_{N+1,t} = \operatorname{Sigmoid}(\mathbf{W}_{N+1}^l \bar{\mathbf{a}}_t^l) \label{equ:mixture-4} \\
\vspace{-2mm}
s_{i,t} = \operatorname{Softmax}_i(\mathbf{W}_{i}^l \bar{\mathbf{a}}_t^l) \label{equ:mixture-2}
\end{gather} 
\end{minipage} }

Here, ${W}_i^l \in {R}^{1 \times D}$ are trainable parameters, $N$ is the number of non-shared experts, and $K$ denotes the activated non-shared experts per MoE layer.

FTS-Text-MoE boosts efficiency by activating only a subset of its 113M parameters, with just 50M active during operation. This selective activation enables the model to utilize specialized experts, boosting efficiency while preserving scalability.

\subsection{Multi-resolution Prediction}

Unlike existing foundational models, which typically output predictions at a fixed horizon, our model incorporates a multi-resolution prediction head followed by \cite{shi2024time}. This prediction head consists of $P$ ($P=4$) output projections, each targeting a specific forecasting horizon $p_j \in \{1,8,32,64\}$, predicting $p_j$ future timesteps. During inference, we apply a greedy scheduling algorithm to concatenate these projections, enabling flexible, arbitrary-length forecasting and enhancing the adaptability of FTS-Text-MoE.

\subsection{Loss Function}
The multi-resolution prediction head comprises several single-layer FFN output projections, each dedicated to a specific forecasting horizon. The final loss combines autoregressive losses across multiple resolutions and an auxiliary balancing loss, as detailed in Appendix \ref{Loss Function}:
\begin{equation}
\begin{aligned}
\mathcal{L} = & \frac{1}{P} \sum_{j=1}^{P} 
\mathcal{L}_{\text{ar}}(\mathbf{A}_{t+1:t+p_j}, \hat{\mathbf{A}}_{t+1:t+p_j}) \\
& + \alpha \mathcal{L}_{\text{aux}},
\end{aligned}
\end{equation}

Where $P$ denotes the number of multi-resolution projections, and $p_j$ represents the horizon length of the $j$-th projection. $\mathcal{L}{\text{ar}}$ (autoregressive loss) quantifies the difference between actual and predicted values to optimize prediction accuracy, while $\mathcal{L}{\text{aux}}$ (auxiliary loss) balances expert utilization, ensuring more even distribution and enhancing training efficiency.

\begin{figure*}[t]
  \centering
  \begin{minipage}[t]{0.30\textwidth}
    \centering
    \includegraphics[width=\textwidth]{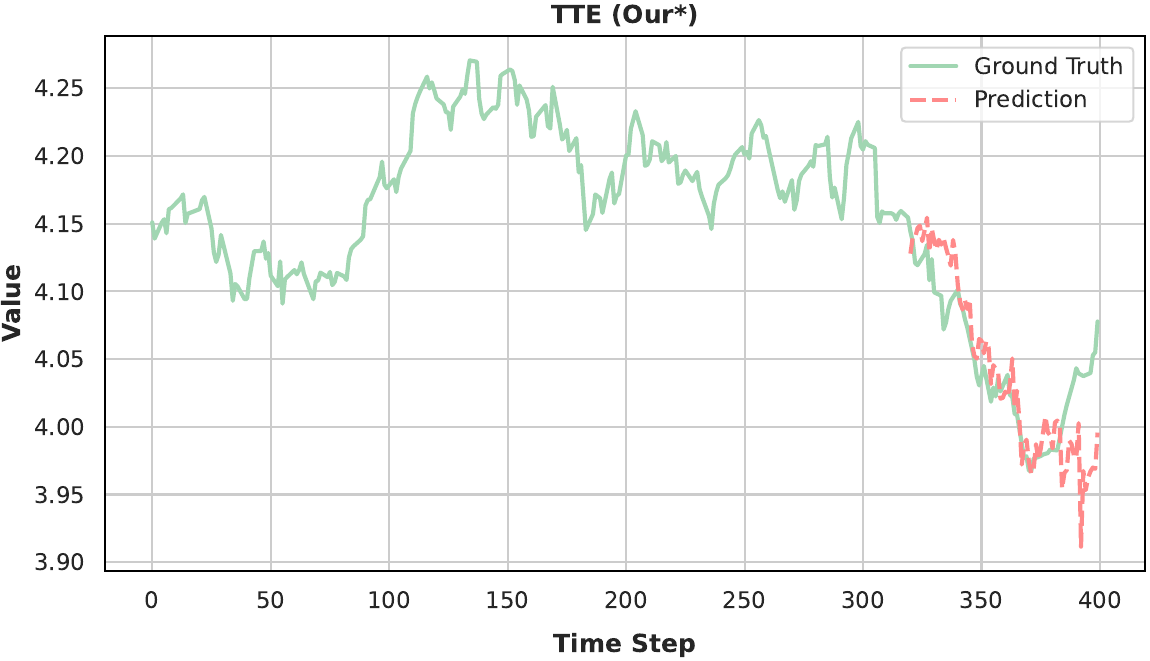}
    \label{fig:ts1}
  \end{minipage}
  \hfill
  \begin{minipage}[t]{0.32\textwidth}
    \centering
    \includegraphics[width=\textwidth]{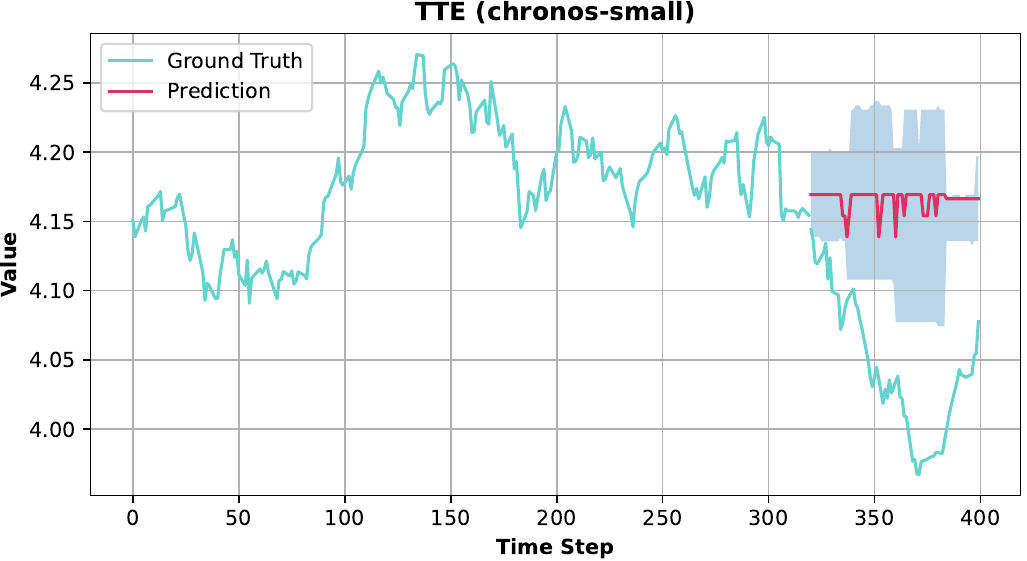}
    \label{fig:ts2}
  \end{minipage}
  \hfill
  \begin{minipage}[t]{0.32\textwidth}
    \centering
    \includegraphics[width=\textwidth]{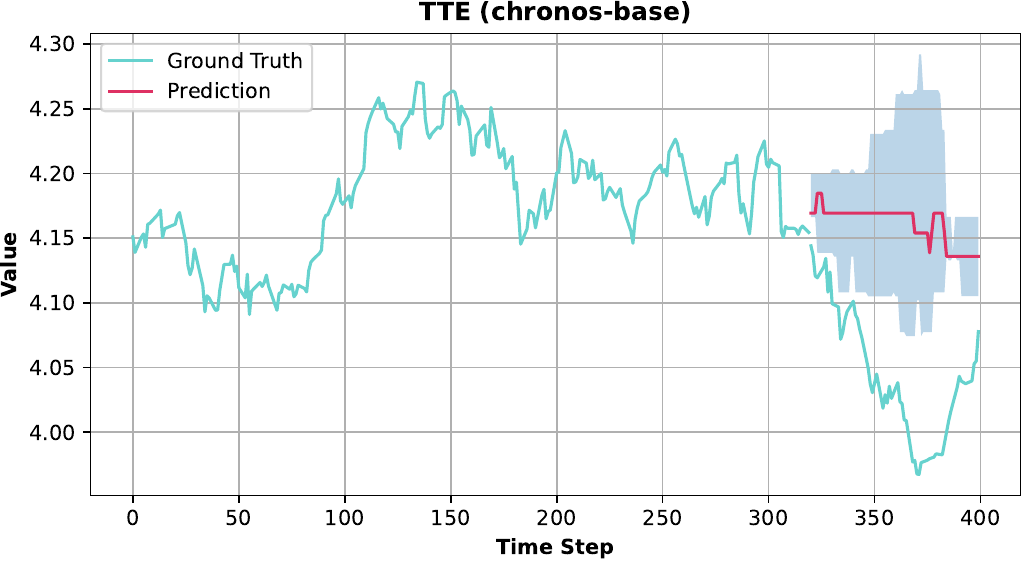}
    \label{fig:ts3}
  \end{minipage}
  
  \vspace{-5mm} 
  \begin{minipage}[t]{0.32\textwidth}
    \centering
    \includegraphics[width=\textwidth]{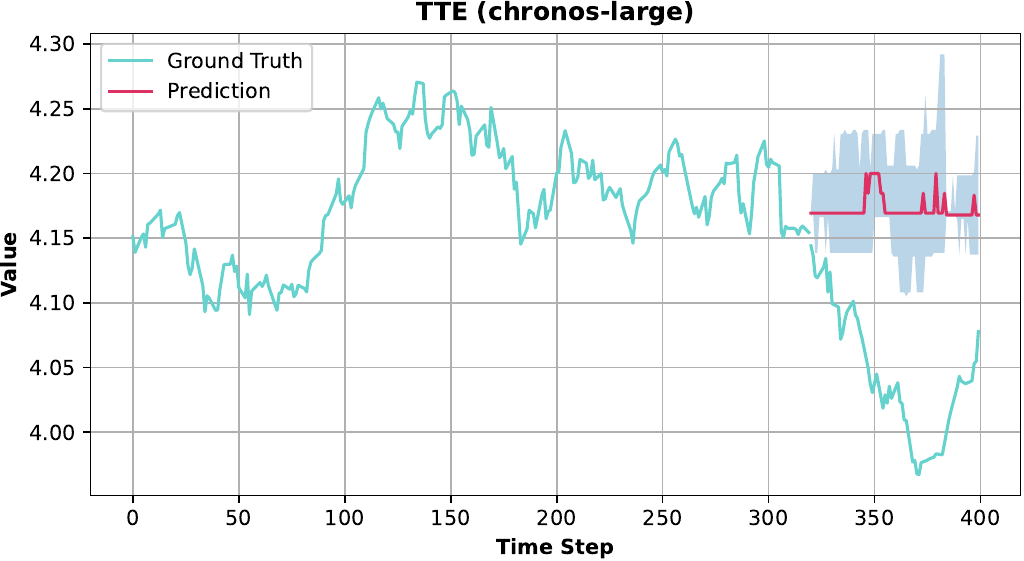}
    \label{fig:ts4}
  \end{minipage}
  \hfill
  \begin{minipage}[t]{0.32\textwidth}
    \centering
    \includegraphics[width=\textwidth]{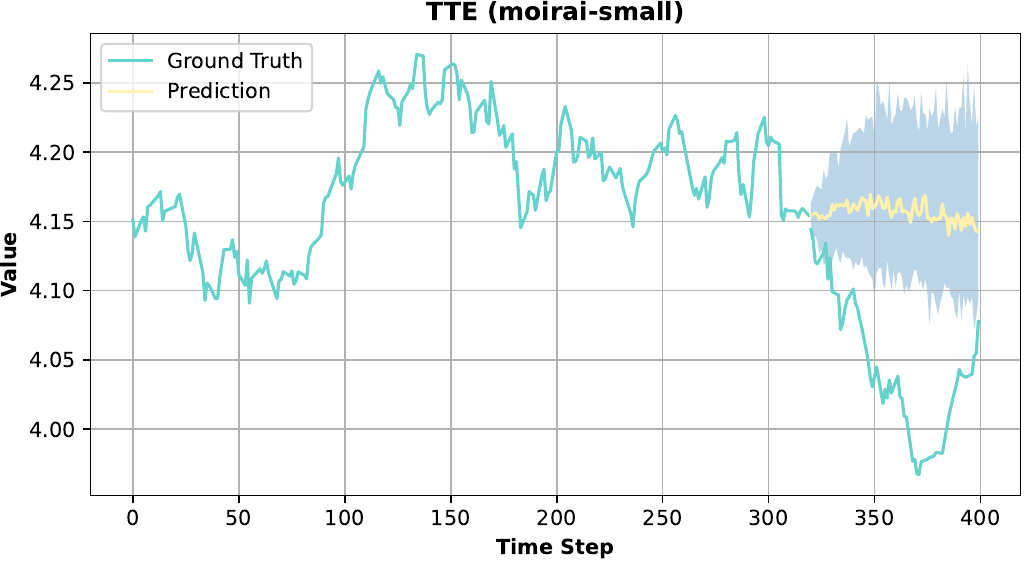}
    \label{fig:ts5}
  \end{minipage}
  \hfill
  \begin{minipage}[t]{0.32\textwidth}
    \centering
    \includegraphics[width=\textwidth]{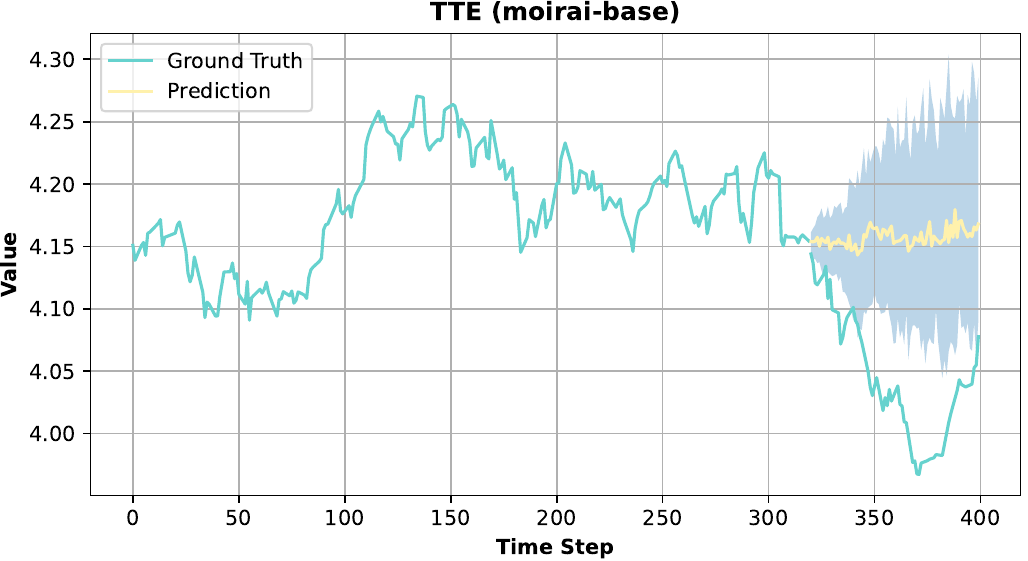}
    \label{fig:ts6}
  \end{minipage}
  \vspace{-8mm}
  \caption{Forecasting results of different models on the TTE stock time series with a forecast horizon of 80 and a context length of 320. The green/blue lines represent the ground truth, while the red/yellow lines indicate the model predictions.}
  \label{fig:TTE_6fig}
\end{figure*}

\begin{figure*}[t]
  \centering
  \begin{subfigure}[b]{0.30\textwidth}
    \centering
    \includegraphics[width=\textwidth]{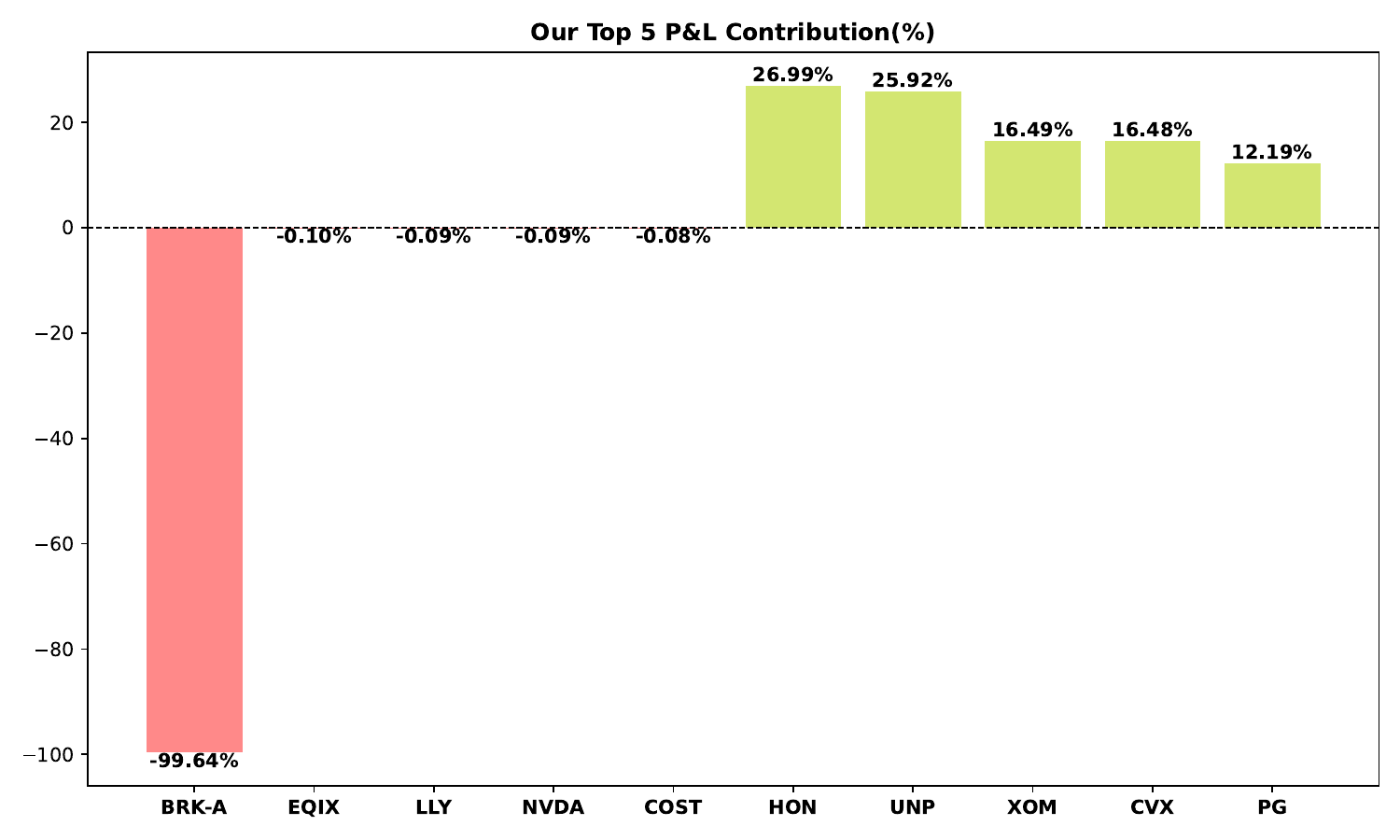}
    \label{fig:bar1}
  \end{subfigure}
  \hfill
  \begin{subfigure}[b]{0.32\textwidth}
    \centering
    \includegraphics[width=\textwidth]{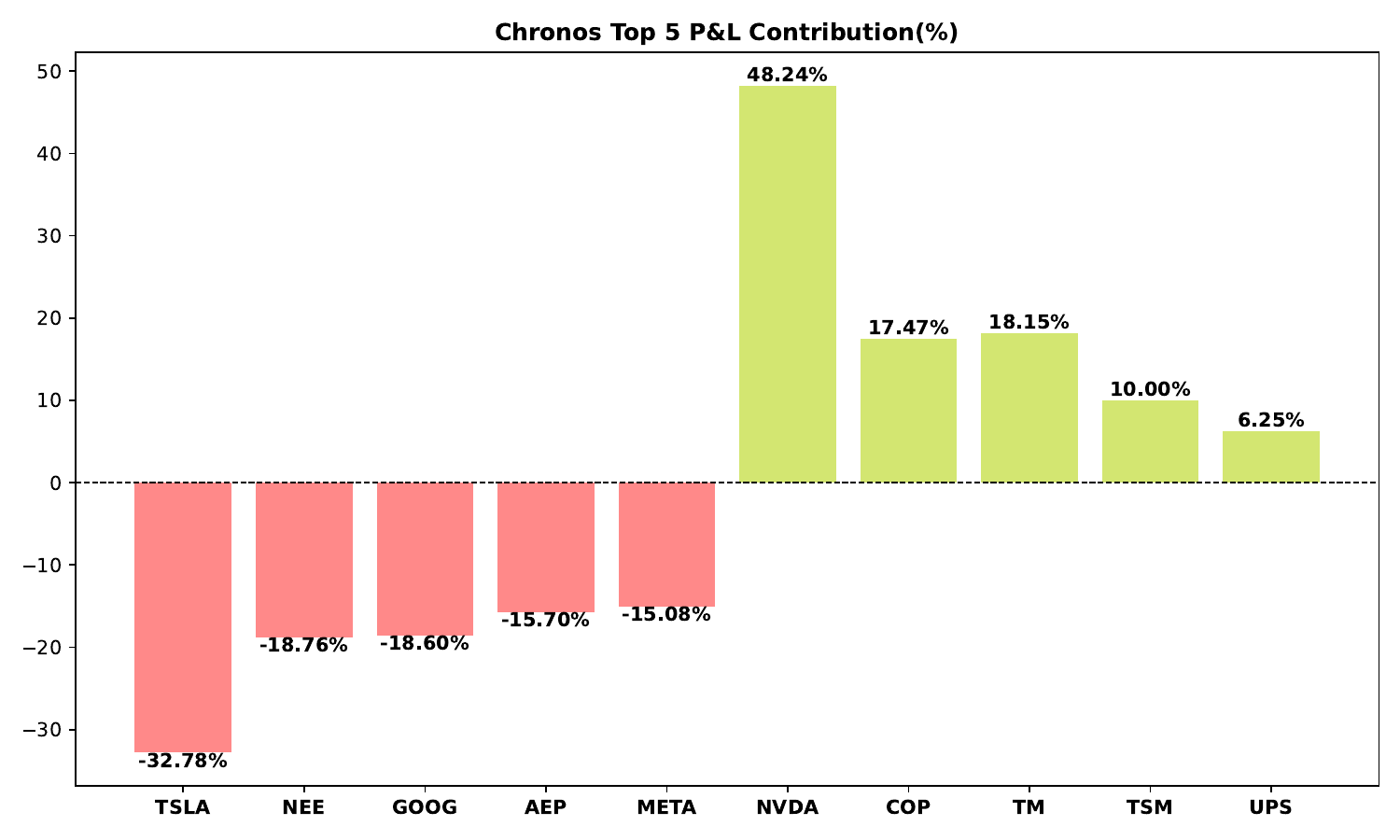}
    \label{fig:bar2}
  \end{subfigure}
  \hfill
  \begin{subfigure}[b]{0.32\textwidth}
    \centering
    \includegraphics[width=\textwidth]{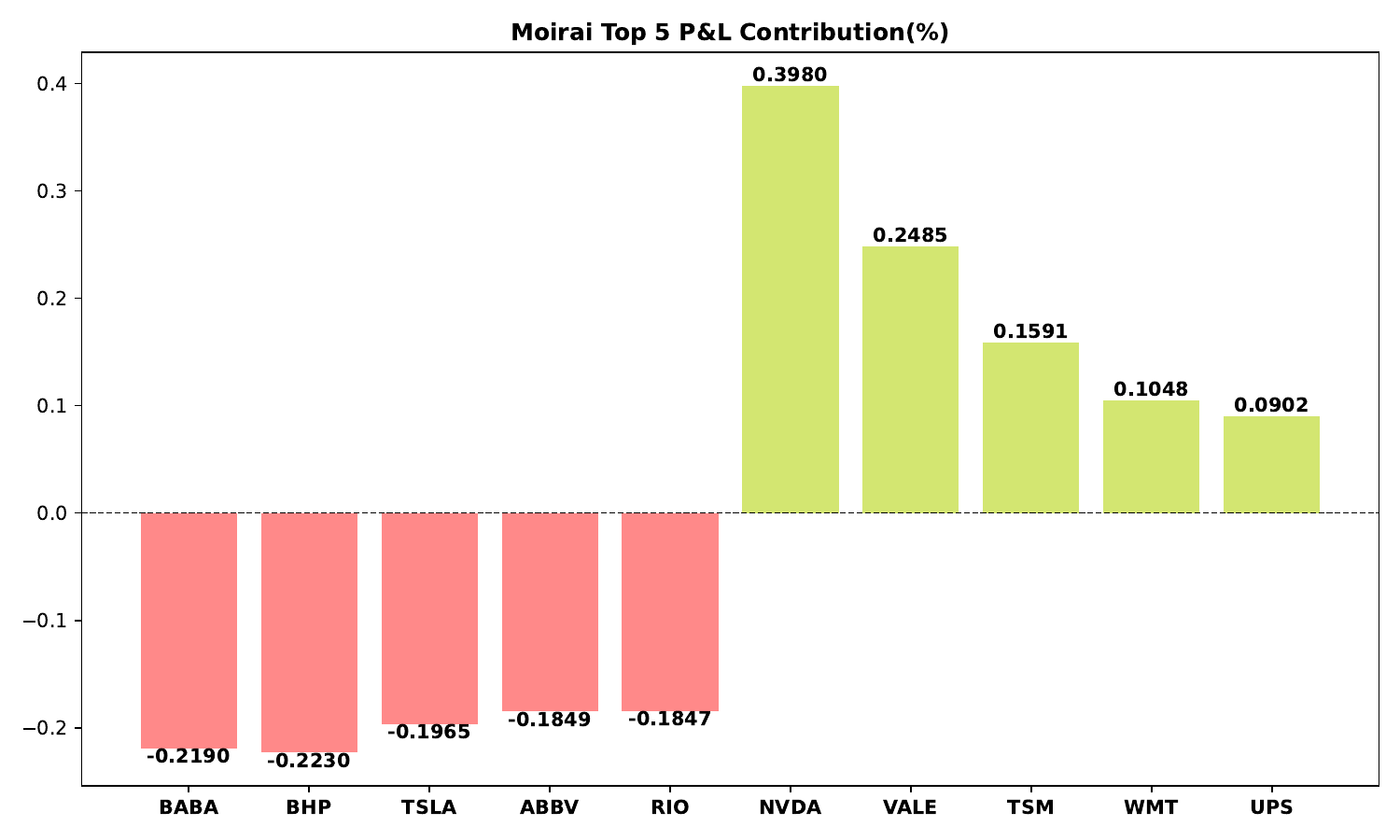}
    \label{fig:bar3}
  \end{subfigure}
   \vspace{-7mm}
  \caption{The top 5 gainers and losers under the three methods. The y-axis represents the proportion of contributions to returns and losses, while the x-axis denotes individual stock.}
  \label{fig:loss_profit}
\end{figure*}

\begin{figure*}[t]
  \centering
  \begin{subfigure}[b]{0.32\textwidth} 
    \centering
\includegraphics[width=\textwidth]{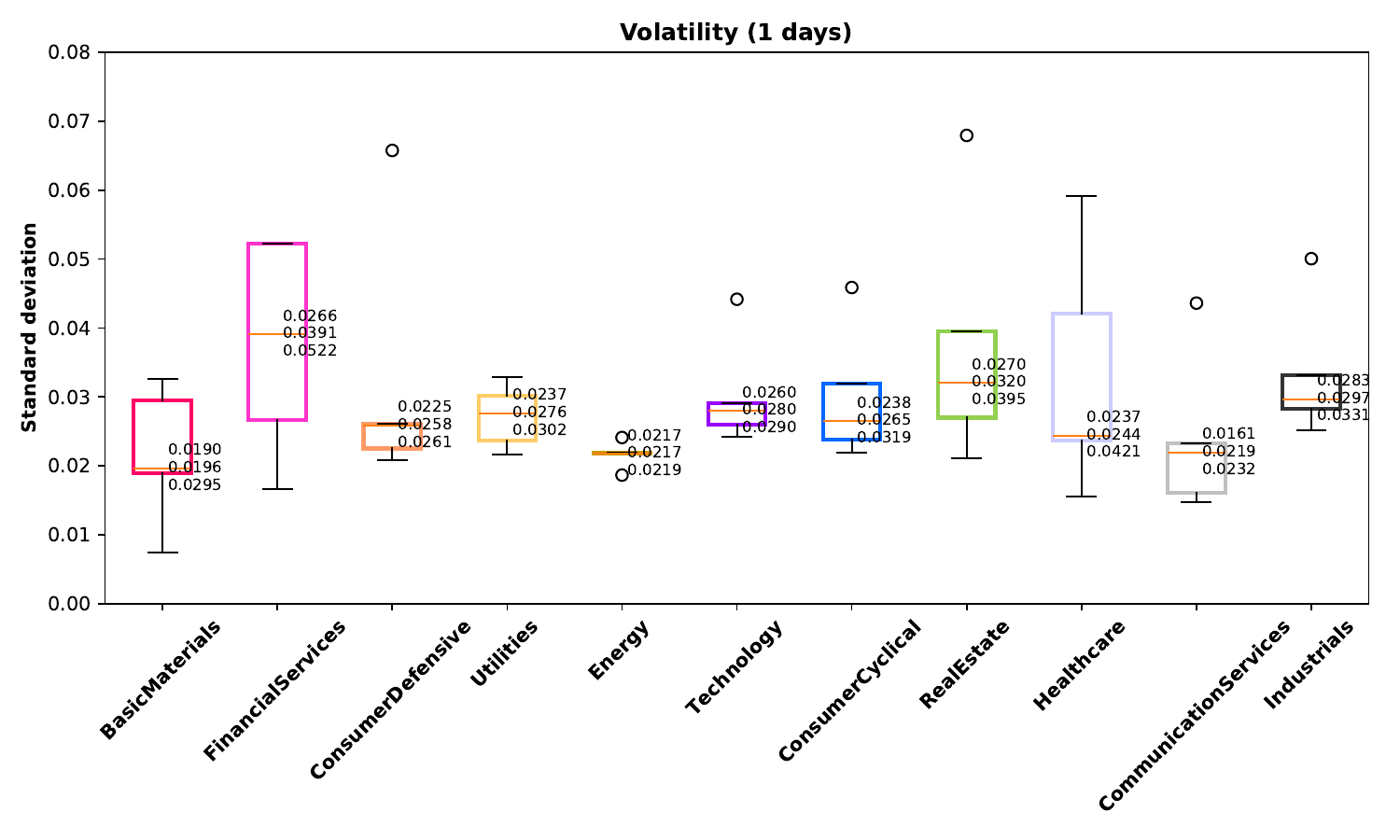}
    \label{fig:box1}
  \end{subfigure}
  \hfill
  \begin{subfigure}[b]{0.32\textwidth}
    \centering
    \includegraphics[width=\textwidth]{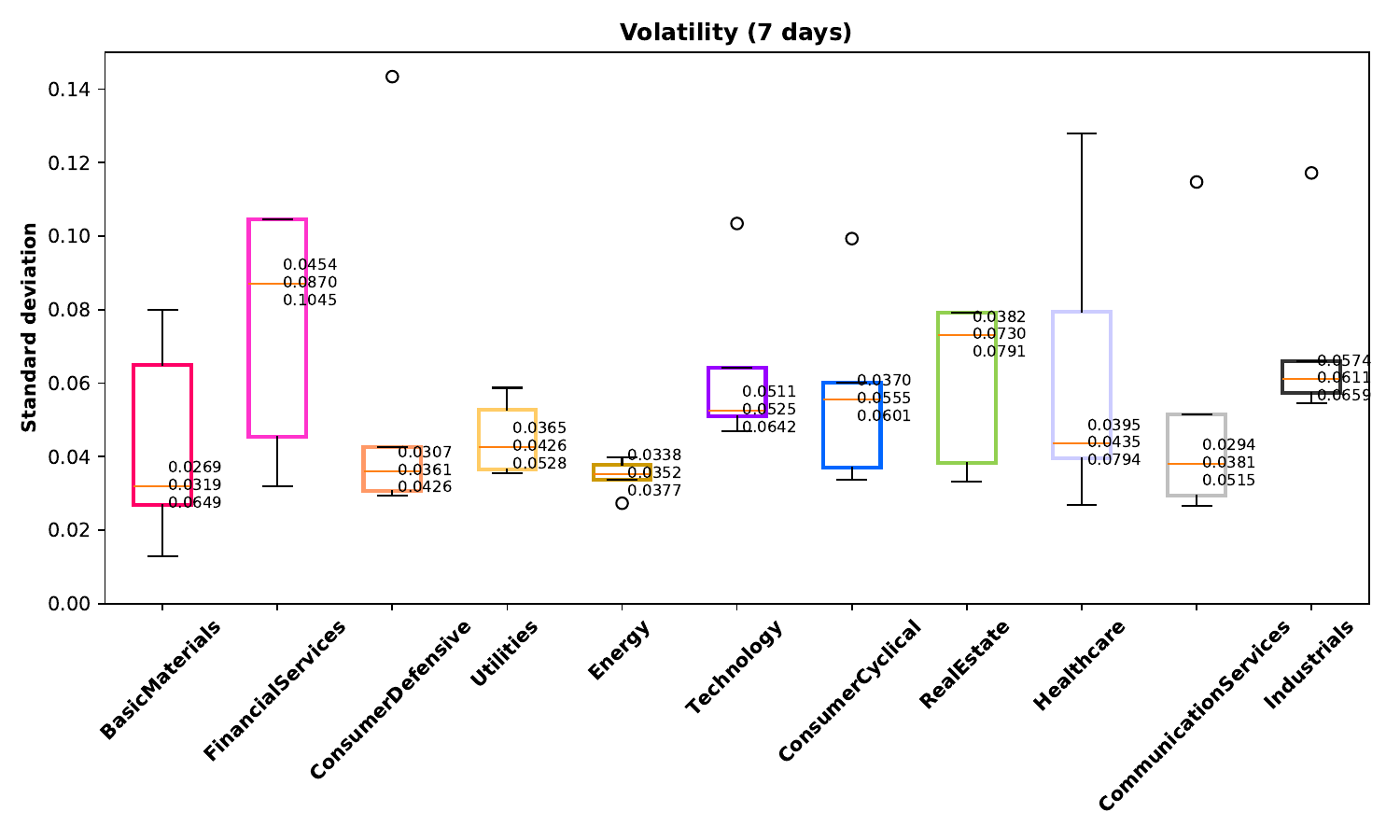}
    \label{fig:box2}
  \end{subfigure}
  \hfill
  \begin{subfigure}[b]{0.32\textwidth}
    \centering
    \includegraphics[width=\textwidth]{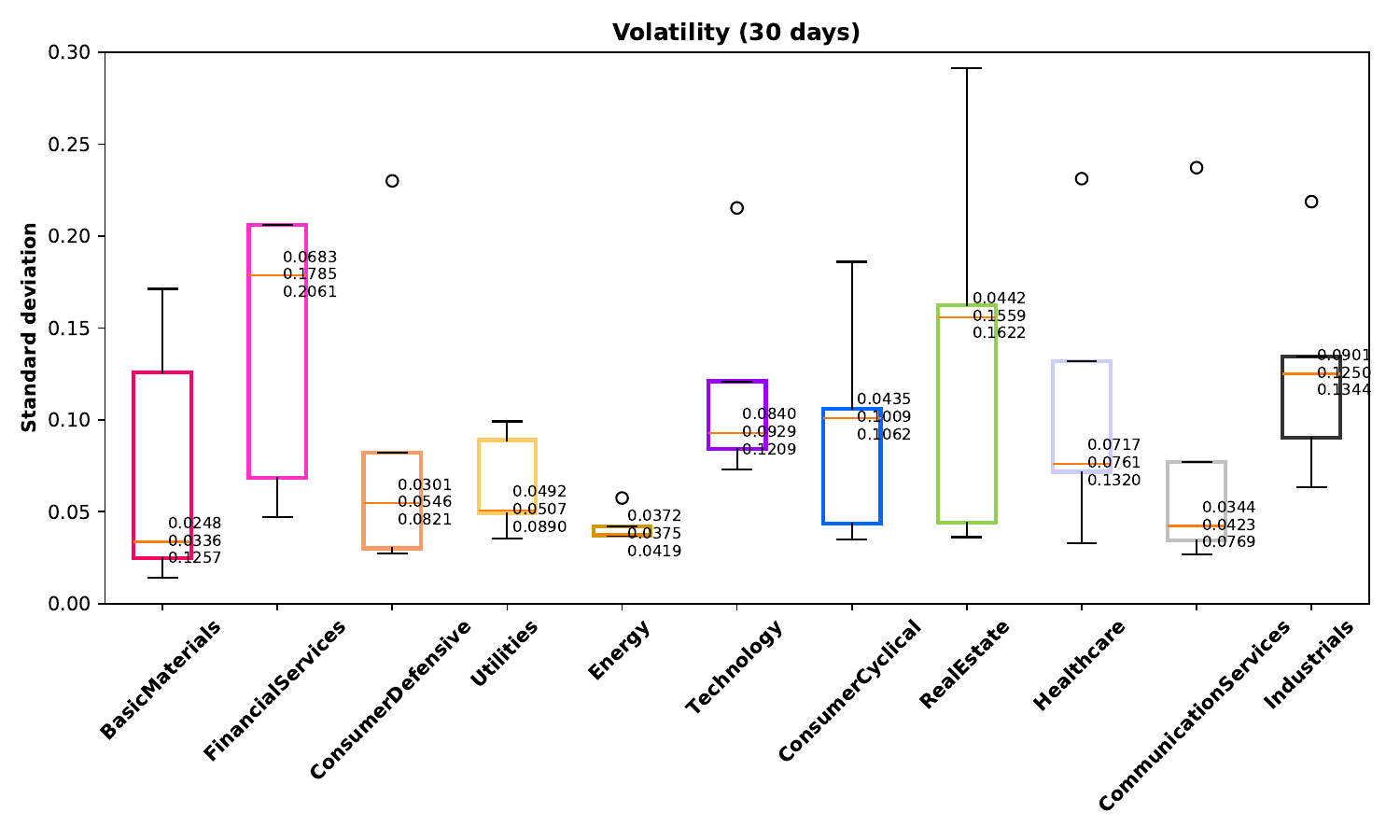}
    \label{fig:box3}
  \end{subfigure}
  \vspace{-7mm}
  \caption{Stock price volatility across 11 industry sectors over 1-day, 7-day, and 30-day sampling periods.}
  \label{fig:three_box}
\end{figure*}

\section{Experiments}
To evaluate the effectiveness of FTS-Text-MoE, we compared it with state-of-the-art language models designed for time series prediction. Details on datasets and metrics can be found in Appendix \ref{Data Sources and Metrics}.

\subsection{Baselines}
We use state-of-the-art models designed for long-term time series forecasting as baselines.

\textbf{Chronos:} Chronos \cite{ansarichronos} discretizes time series into intervals by scaling and quantizing real values. This enables training language models on the "language of time series" without altering the model architecture. We evaluate Chronos models (t5-small\footnote{\scriptsize\url{https://huggingface.co/amazon/chronos-t5-small}}/base\footnote{\scriptsize\url{https://huggingface.co/amazon/chronos-t5-base}}/large\footnote{\scriptsize\url{https://huggingface.co/amazon/chronos-t5-large}}).

\textbf{Moirai:} Moirai \cite{woo2024unified} concatenates multivariate time series into a single sequence with [MASK] tokens for prediction, and the model uses multi-granularity patching to handle different frequencies. We evaluate Moirai models (t5-small\footnote{\scriptsize\url{https://huggingface.co/Salesforce/moirai-moe-1.0-R-small}}/base\footnote{\scriptsize\url{https://huggingface.co/Salesforce/moirai-moe-1.0-R-base}}).

\subsection{Performance Comparison}
In this section, we quantitatively and qualitatively assess the predictive performance of our FTS-Text-MoE model against those baselines. 

\subsubsection{Effect of Text Source and Length on Prediction Accuracy}

We compare performance using Mean Squared Error (MSE) and Mean Absolute Error (MAE) across 11 industry sectors, integrating stock time series data with various textual inputs—summarized news and summarized X comments. Lower MSE and MAE values indicate superior predictive accuracy.

Our analysis highlights that adding summarized news or tweet inputs significantly improves stock movement prediction accuracy (Table \ref{tab:different-method-comparison}). Specifically, using summarized X comments results in consistently lower MSE and MAE compared to models using only news or other inputs. This indicates that distilled sentiment signals effectively reduce noise and enhance prediction accuracy. Industry-level analysis reveals that sectors like Healthcare, Consumer Cyclical, and Communication Services exhibit lower prediction errors, suggesting a strong reliance on textual signals and improved performance with text integration. In contrast, industries like Energy and Industrials, which experience higher volatility, show improved accuracy with textual inputs but still exhibit larger errors.

\subsubsection{Prediction Performance of TS Models}

When compared to the baseline models, Chronos and Moirai, FTS-Text-MoE shows improved performance with the addition of text inputs (such as news and tweets), but still falls short of surpassing Chronos and Moirai in some industries and scenarios (Table \ref{tab:different-method-comparison}). Chronos and Moirai models exhibit a more conservative prediction approach, especially under high uncertainty and market volatility, smoothing the curve to reduce errors and maintain lower MSE and MAE. In contrast, FTS-Text-MoE adopts a more proactive strategy, capturing short-term volatility, which benefits quick market shifts but increases prediction errors during periods of significant fluctuations\ref{fig:TTE_6fig}.

\begin{table}[t]
  \centering
  \resizebox{1\linewidth}{!}{ 
    \begin{tabular}{lccc}
      \hline
      \textbf{Approach} & \textbf{Overall ($\uparrow$)} & \textbf{Std. Dev. ($\downarrow$)} & \textbf{Sharpe ($\uparrow$)} \\
      \hline
      1/N & 0.0570 & 5.132e-3 & 0.1725 \\
      S\&P500 Index & -0.0082 & 5.459e-3 & -0.3002 \\
      Moirai & 0.0079 & 1.741e-3 & 0.3320 \\
      Chronos & -0.0077 & \textbf{1.288e-3} & 0.1223 \\
      Our & \textbf{0.1347} & 5.290e-3 & \textbf{1.0818} \\
      \hline
    \end{tabular}
  }
  \caption{Performance comparison of different portfolio approaches. Overall, Std. Dev., and Sharpe can be referred to in Appendix \ref{Evaluation Metrics}, formulas \ref{Overall}, \ref{Std. Dev.} and \ref{Sharpe}.}
  \label{tab:portfolio_performance}
\end{table}

\subsection{Robustness Analysis of FTS-Text-MoE}
We analyzed stock price volatility across sectors (see Figure \ref{fig:three_box}), measuring it as the standard deviation of changes over a given sampling period ($P=1,7,30$):
Higher volatility indicates greater risk, which generally increases with a longer sampling period $P$.
\vspace{-3.8mm}
\small \begin{equation} \sigma_{P} = \sqrt{\frac{1}{N_{P}}\sum_{i=d_{0}}^{d_{n}-P}(v_{i}^{P}-\overline{v_{P}})^{2}}, \end{equation} \normalsize
where $d_{0}$ and $d_{n}$ are the first and last days, $v_{i}^{P}=y_{i+P}-y_{i}$ represents the value change over $P$, and $\overline{v_{P}}$ is the average change:
\vspace{-3mm}
\small \begin{equation} \overline{v_{P}}=\frac{1}{N_{P}}\sum_{i=d_{0}}^{d_{n}-P}v_{i}^{P}. \end{equation} \normalsize

\subsection{Comparison of Return and Loss Distributions}
Compared to FTS-Text-MoE, Chronos and Moirai, which rely solely on historical time-series data, adopt a more passive prediction strategy. In contrast, FTS-Text-MoE integrates real-world textual information, resulting in a more aggressive forecasting approach. Figure \ref{fig:loss_profit} shows the top five gainers and losers for FTS-Text-MoE and the baseline models. The return distribution for FTS-Text-MoE is noticeably more uneven, indicating a more aggressive and concentrated investment strategy.

\subsection{Portfolio Optimization}

We evaluate portfolio performance using three key metrics across four baselines:
\vspace{-3mm}
\begin{itemize}  
\setlength{\itemsep}{0pt}  
    \item \textbf{1/N Portfolio}: Equally weighted across 11 stocks.  
    \item \textbf{S\&P 500 Index}: Market benchmark.  
    \item \textbf{Moirai/Chronos Positive Prediction Portfolios}: Invests only in positively predicted stocks, equally weighted.    
\end{itemize}

As shown in Table \ref{tab:portfolio_performance}, FTS-Text-MoE, following the same investment approach, outperforms Moirai/Chronos, particularly in terms of overall returns and Sharpe ratio. With an Overall return of 0.1347, it demonstrates superior cumulative returns, while its Sharpe ratio of 1.0818 indicates better risk-adjusted performance. However, its higher standard deviation (Std. Dev. = 5.290e-3) suggests increased return volatility, likely due to the model’s responsiveness to time series fluctuations, making it more sensitive to short-term market dynamics.

\section{Conclusion}
This study proposes FTS-Text-MoE, which effectively integrates time series and textual data to overcome the input-output length limitations faced by existing large-scale financial forecasting models when processing multi-source data and performing long-sequence predictions. The proposed model employs a sparse Transformer decoder combined with an MoE mechanism, significantly enhancing computational efficiency, and utilizes multi-resolution prediction heads to enable flexible forecasting across different temporal scales.
Experimental results demonstrate that FTS-Text-MoE achieves superior performance in financial time series prediction tasks, effectively integrating quantitative and sentiment data for a more comprehensive market analysis.

\clearpage
\section{Limitations}
The proposed FTS-Text-MoE integrates financial numerical data with textual information and reduces computational complexity using a sparse Transformer decoder, but it has limitations. Firstly, the model aligns textual and time-series data at the same time step, yet news and tweets often have a delayed impact on stock prices. This lag, where market information takes time to propagate, may affect prediction accuracy. Secondly, news and social media may contain false or misleading information, negatively influencing forecasts. Future research will focus on handling lag and improving text authenticity verification to enhance model robustness and accuracy.

\bibliography{custom}

\clearpage

\appendix

\section{Implementation Details}
\subsection{Loss Function}
\label{Loss Function}
Integrating extensive textual information into financial time series data poses a challenge for training stability in FTS-Text-MoE. To enhance robustness against outliers and improve stability, we employ the Huber loss. The autoregressive loss, projected from a single-layer FFN, is defined as follows:

\begin{equation} 
\resizebox{0.9\hsize}{!}{$
\mathcal{L}_{\text{ar}} \left( a_t, \hat{a}_t \right) =
\left\{
\begin{array}{ll}
\frac{1}{2} \left( a_t - \hat{a}_t \right)^{2}, & \text{if } \left| a_t - \hat{a}_t \right| \leq \delta, \\
\delta \times \left( \left| a_t - \hat{a}_t \right| - \frac{1}{2} \times \delta \right), & \text{otherwise}.
\end{array}
\right.
$}
\end{equation}
where $\delta$ is a hyperparameter that balances the L1 and L2 loss components.

In MoE architectures, optimizing purely for accuracy often leads to uneven expert utilization, with the model favoring a small subset of experts, limiting broader training. To address this, inspired by \cite{dai2024deepseekmoe} and \cite{fedus2022switch}, we introduce an auxiliary loss for expert load balancing. This loss aligns the proportion of tokens assigned to each expert ($f_i$) with its routing probability ($r_i$), ensuring a more balanced distribution.
\begin{equation}
\begin{aligned}
\mathcal{L}_{\text{aux}} = N \sum_{i=1}^{N} f_i r_i,\ r_i = \frac{1}{T} \sum_{t=1}^{T} s_{i,t},\quad\quad \\ f_i = \frac{1}{KT} \sum_{t=1}^{T} \mathbb{I} 
\Big(\text{Time step } t \text{ selects Expert } i \Big)
\end{aligned}
\label{equ:aux_loss}
\end{equation}
where $\mathbb{I}$ is an indicator function.


The multi-resolution forecasting heads utilize multiple output projections from a single-layer FFN, each corresponding to a different forecasting horizon. During training, errors across scales are aggregated into a composite loss. The final loss function combines the multi-resolution autoregressive loss and the auxiliary balancing loss:
\begin{equation} 
\mathcal{L} = \frac{1}{P} \sum_{j=1}^{P} \mathcal{L}_{\text{ar}} \Big( \mathbf{A}_{t+1:t+p_{j}}, \hat{\mathbf{A}}_{t+1:t+p_{j}} \Big) + \alpha \mathcal{L}_{\text{aux}},
\end{equation}
where $P$ is the number of multi-resolution projections, and $p_j$ denotes the horizon of the $j$-th projection.

\subsection{Model Details}
FTS-Text-MoE employs a 12-layer architecture with 12 attention heads per layer and integrates an MoE module with 8 experts per layer. For each token, the top-2 experts ($K = 2$) are selected. The model features a hidden dimension ($d_{\text{model}}$) of 384, a feed-forward network dimension ($d_{f}$) of 1536, and each expert has a dimension of 192 ($d_{\text{expert}}$). This design results in 50M active parameters during inference, with a total parameter count of 113M.

\subsection{Training Configuration}
Each model is trained for 10,000 steps with a batch size of 64 and a maximum sequence length of 1,024. Each iteration processes 64 time steps. The output projection employs prediction horizons of \{1, 8, 32, 64\}, and the auxiliary loss coefficient $\alpha$ is set to 0.02. Optimization is performed using the AdamW optimizer with hyperparameters: learning rate $lr = 5\text{e-}5$, weight decay $\lambda = 0.1$, $\beta_1 = 0.9$, and $\beta_2 = 0.95$. The learning rate follows a schedule with a linear warm-up for the first 10,000 steps, after which cosine annealing is applied. Training is conducted on two NVIDIA A6000-48G GPUs with FP16 precision. To improve batch efficiency and handle variable-length sequences, we employ sequence packing to minimize padding overhead.

\section{Data Sources and Metrics}
\label{Data Sources and Metrics}

\subsection{Time Series Data}
Stock price data from Yahoo Finance includes 55 stocks across 11 industries with daily open, high, low, close prices, and trading volume. The adjusted closing price is used as the primary measure of stock value over 20 years, yielding 191,512 sample points. See Table~\ref{tab:stock_data} for details.

\begin{table*}[t]
    \centering
    \renewcommand{\arraystretch}{0.88} 
    \setlength{\tabcolsep}{10pt} 
    \resizebox{\textwidth}{!}{ 
    \begin{tabular}{|>{\centering\arraybackslash}m{4cm}|
                        >{\centering\arraybackslash}m{2cm}|
                        >{\centering\arraybackslash}m{1.5cm}|
                        >{\centering\arraybackslash}m{3cm}|
                        >{\centering\arraybackslash}m{3cm}|}
        \hline
        \textbf{Sector} & \textbf{Stock Symbol} & \textbf{Total Days} & \textbf{First Date} & \textbf{Last Date} \\
        \hline
        \multirow{5}{*}{Basic Materials} & BHP & 5568 & 2009/10/21 & 2025/1/17 \\
        \cline{2-5}
        & RIO & 4461 & 2012/10/31 & 2025/1/16 \\
        \cline{2-5}
        & SHW & 3390 & 2015/10/6 & 2025/1/15 \\
        \cline{2-5}
        & VALE & 5465 & 2010/2/1 & 2025/1/17 \\
        \cline{2-5}
        & APD & 4796 & 2011/12/2 & 2025/1/17 \\
        \hline
        \multirow{5}{*}{Financial Services} & BRK-A & 5570 & 2009/10/19 & 2025/1/18 \\
        \cline{2-5}
        & V & 5392 & 2010/4/15 & 2025/1/19 \\
        \cline{2-5}
        & JPM & 1840 & 2020/1/5 & 2025/1/17 \\
        \cline{2-5}
        & MA & 1840 & 2020/1/5 & 2025/1/17 \\
        \cline{2-5}
        & BAC & 1840 & 2020/1/5 & 2025/1/18 \\
        \hline
        \multirow{5}{*}{Consumer Defensive} & WMT & 2240 & 2018/12/1 & 2025/1/19 \\
        \cline{2-5}
        & PG & 1840 & 2020/1/5 & 2025/1/18 \\
        \cline{2-5}
        & KO & 5575 & 2009/10/14 & 2025/1/18 \\
        \cline{2-5}
        & PEP & 5548 & 2009/11/10 & 2025/1/19 \\
        \cline{2-5}
        & COST & 5449 & 2010/2/17 & 2025/1/18 \\
        \hline
        \multirow{5}{*}{Utilities} & NEE & 1639 & 2020/7/24 & 2025/1/17 \\
        \cline{2-5}
        & DUK & 5381 & 2010/4/26 & 2025/1/17 \\
        \cline{2-5}
        & D & 5387 & 2010/4/20 & 2025/1/18 \\
        \cline{2-5}
        & SO & 5324 & 2010/6/22 & 2025/1/18 \\
        \cline{2-5}
        & AEP & 5579 & 2009/10/9 & 2025/1/16 \\
        \hline
        \multirow{5}{*}{Energy} & XOM & 1973 & 2019/8/25 & 2025/1/18 \\
        \cline{2-5}
        & CVX & 3090 & 2016/8/3 & 2025/1/18 \\
        \cline{2-5}
        & SHEL & 1840 & 2020/1/5 & 2025/1/17 \\
        \cline{2-5}
        & TTE & 1839 & 2020/1/5 & 2025/1/16 \\
        \cline{2-5}
        & COP & 5541 & 2009/11/17 & 2025/1/17 \\
        \hline
        \multirow{5}{*}{Technology} & AAPL & 1844 & 2020/1/1 & 2025/1/18 \\
        \cline{2-5}
        & MSFT & 1840 & 2020/1/5 & 2025/1/18 \\
        \cline{2-5}
        & TSM & 5392 & 2010/4/15 & 2025/1/19 \\
        \cline{2-5}
        & NVDA & 1615 & 2020/8/17 & 2025/1/20 \\
        \cline{2-5}
        & AVGO & 1839 & 2020/1/6 & 2025/1/19 \\
        \hline
        \multirow{5}{*}{Consumer Cyclical} & AMZN & 1844 & 2020/1/1 & 2025/1/19 \\
        \cline{2-5}
        & TSLA & 1841 & 2020/1/4 & 2025/1/19 \\
        \cline{2-5}
        & HD & 1841 & 2020/1/4 & 2025/1/17 \\
        \cline{2-5}
        & BABA & 3725 & 2014/11/7 & 2025/1/17 \\
        \cline{2-5}
        & TM & 5465 & 2010/2/1 & 2025/1/18 \\
        \hline
        \multirow{5}{*}{Real Estate} & AMT & 5526 & 2009/12/2 & 2025/1/17 \\
        \cline{2-5}
        & PLD & 5322 & 2010/6/24 & 2025/1/18 \\
        \cline{2-5}
        & CCI & 5442 & 2010/2/23 & 2025/1/16 \\
        \cline{2-5}
        & EQIX & 3469 & 2015/7/21 & 2025/1/17 \\
        \cline{2-5}
        & PSA & 3286 & 2016/1/20 & 2025/1/17 \\
        \hline
        \multirow{5}{*}{Healthcare} & UNH & 1839 & 2020/1/6 & 2025/1/20 \\
        \cline{2-5}
        & JNJ & 1836 & 2020/1/9 & 2025/1/18 \\
        \cline{2-5}
        & LLY & 1840 & 2020/1/5 & 2025/1/18 \\
        \cline{2-5}
        & PFE & 1839 & 2020/1/6 & 2025/1/18 \\
        \cline{2-5}
        & ABBV & 4398 & 2013/1/3 & 2025/1/17 \\
        \hline
        \multirow{5}{*}{Communication Services} & GOOG & 2173 & 2019/2/6 & 2025/1/19 \\
        \cline{2-5}
        & META & 1841 & 2020/1/4 & 2025/1/19 \\
        \cline{2-5}
        & VZ & 1833 & 2020/1/12 & 2025/1/19 \\
        \cline{2-5}
        & CMCSA & 2844 & 2017/4/6 & 2025/1/17 \\
        \cline{2-5}
        & DIS & 2046 & 2019/6/13 & 2025/1/18 \\
        \hline
        \multirow{5}{*}{Industrials} & UPS & 5538 & 2009/11/20 & 2025/1/18 \\
        \cline{2-5}
        & UNP & 5555 & 2009/11/3 & 2025/1/18 \\
        \cline{2-5}
        & HON & 1837 & 2020/1/8 & 2025/1/18 \\
        \cline{2-5}
        & LMT & 1841 & 2020/1/4 & 2025/1/18 \\
        \cline{2-5}
        & CAT & 5554 & 2009/11/4 & 2025/1/17 \\
        \hline
    \end{tabular}
    }
    \caption{Stock Price Series: Collection Periods and Total Days by Sector}
    \label{tab:stock_data}
\end{table*}

\subsection{Tweet Data}
Given the vast number of daily tweets, we use a dataset clustered via the BERTopic pipeline. It includes tweets about the top 5 stocks in 11 sectors from 2020 to 2022, collected via the Twitter API. The dataset comprises 637,395 tweets. See Table~\ref{tab:stock_tweets} for details.

\begin{table*}[t]
    \centering
    \renewcommand{\arraystretch}{0.88} 
    \setlength{\tabcolsep}{10pt} 
    \resizebox{\textwidth}{!}{
    \begin{tabular}{|>{\centering\arraybackslash}m{2cm}|
                        >{\centering\arraybackslash}m{2cm}|
                        >{\centering\arraybackslash}m{2cm}|
                        >{\centering\arraybackslash}m{3.5cm}|
                        >{\centering\arraybackslash}m{3.5cm}|}
        \hline
        \textbf{Company} & \textbf{Total Days} & \textbf{Daily Tweet Count} & \textbf{First Date} & \textbf{Last Date} \\
        \hline
        AAPL & 1090 & 83964 & 2020/1/1 & 2022/12/31 \\
        \hline
        ABBV & 914 & 3200 & 2020/1/4 & 2022/12/21 \\
        \hline
        AEP & 416 & 2156 & 2020/1/12 & 2022/12/28 \\
        \hline
        AMT & 546 & 2536 & 2020/1/10 & 2022/12/24 \\
        \hline
        AMZN & 1090 & 58193 & 2020/1/1 & 2022/12/31 \\
        \hline
        APD & 444 & 2283 & 2020/1/4 & 2022/12/16 \\
        \hline
        AVGO & 776 & 2466 & 2020/1/6 & 2022/12/24 \\
        \hline
        BABA & 978 & 18761 & 2020/1/4 & 2022/12/24 \\
        \hline
        BAC & 1046 & 7353 & 2020/1/5 & 2022/12/24 \\
        \hline
        BHP & 446 & 1433 & 2020/1/7 & 2022/12/23 \\
        \hline
        BRK-A & 414 & 1910 & 2020/1/4 & 2022/12/29 \\
        \hline
        CAT & 937 & 4823 & 2020/1/6 & 2022/12/24 \\
        \hline
        CCI & 339 & 1651 & 2020/1/5 & 2022/12/24 \\
        \hline
        CMCSA & 806 & 2412 & 2020/1/6 & 2022/12/24 \\
        \hline
        COP & 689 & 2027 & 2020/1/19 & 2022/12/24 \\
        \hline
        COST & 1003 & 4780 & 2020/1/7 & 2022/12/24 \\
        \hline
        CVX & 824 & 6907 & 2020/1/5 & 2022/12/24 \\
        \hline
        D & 462 & 1353 & 2020/1/12 & 2022/12/24 \\
        \hline
        DIS & 1064 & 14146 & 2020/1/8 & 2022/12/24 \\
        \hline
        DUK & 357 & 1688 & 2020/1/9 & 2022/12/24 \\
        \hline
        EQIX & 441 & 2505 & 2020/1/5 & 2022/12/28 \\
        \hline
        GOOG & 1085 & 12539 & 2020/1/4 & 2022/12/24 \\
        \hline
        HD & 923 & 4837 & 2020/1/4 & 2022/12/24 \\
        \hline
        HON & 464 & 1436 & 2020/1/8 & 2022/12/24 \\
        \hline
        JNJ & 884 & 6104 & 2020/1/9 & 2022/12/20 \\
        \hline
        JPM & 950 & 7783 & 2020/1/5 & 2022/12/24 \\
        \hline
        KO & 828 & 3915 & 2020/1/7 & 2022/12/24 \\
        \hline
        LLY & 745 & 2781 & 2020/1/5 & 2022/12/24 \\
        \hline
        LMT & 733 & 2620 & 2020/1/4 & 2022/10/17 \\
        \hline
        MA & 913 & 3408 & 2020/1/5 & 2022/12/24 \\
        \hline
        META & 965 & 44182 & 2020/1/4 & 2022/12/24 \\
        \hline
        MSFT & 963 & 31153 & 2020/1/4 & 2022/12/24 \\
        \hline
        NEE & 559 & 1637 & 2020/1/4 & 2022/12/24 \\
        \hline
        PEP & 669 & 2196 & 2020/1/15 & 2022/12/24 \\
        \hline
        PFE & 1031 & 12021 & 2020/1/6 & 2022/12/24 \\
        \hline
        PG & 727 & 2665 & 2020/1/5 & 2022/12/24 \\
        \hline
        PLD & 505 & 2362 & 2020/1/9 & 2022/12/24 \\
        \hline
        PSA & 499 & 2482 & 2020/1/11 & 2022/12/27 \\
        \hline
        RIO & 578 & 2005 & 2020/1/6 & 2022/12/24 \\
        \hline
        SHEL & 406 & 1688 & 2020/1/5 & 2022/12/23 \\
        \hline
        SHW & 398 & 1939 & 2020/1/4 & 2022/12/23 \\
        \hline
        SO & 365 & 1590 & 2020/1/4 & 2022/12/29 \\
        \hline
        TM & 560 & 2053 & 2020/1/5 & 2022/12/23 \\
        \hline
        TSLA & 973 & 181368 & 2020/1/4 & 2022/12/24 \\
        \hline
        TSM & 786 & 3850 & 2020/1/6 & 2022/12/24 \\
        \hline
        TTE & 404 & 1911 & 2020/1/5 & 2022/12/24 \\
        \hline
        UNH & 818 & 2550 & 2020/1/6 & 2022/12/24 \\
        \hline
        UNP & 398 & 1557 & 2020/1/4 & 2022/12/22 \\
        \hline
        UPS & 803 & 2805 & 2020/1/4 & 2022/12/23 \\
        \hline
        V & 936 & 5293 & 2020/1/4 & 2022/12/24 \\
        \hline
        VALE & 660 & 1937 & 2020/1/4 & 2022/12/24 \\
        \hline
        VZ & 745 & 3119 & 2020/1/12 & 2022/12/24 \\
        \hline
        WMT & 1076 & 11164 & 2020/1/4 & 2022/12/24 \\
        \hline
        XOM & 1060 & 7933 & 2020/1/5 & 2022/12/24 \\
        \hline
        NVDA & 1074 & 33965 & 2020/1/4 & 2022/12/24 \\
        \hline
    \end{tabular}
    }
    \caption{Stock Tweets: Collection Periods and Total Days by Sector.}
    \label{tab:stock_tweets}
\end{table*}

\subsection{News Data}
\label{News Data}
We extract news data from the FNSPID dataset, focusing on articles related to 55 S\&P 500 stocks listed on NASDAQ from 1999 to 2023. This dataset includes URLs, headlines, and full news texts.

To ensure up-to-date coverage, we addressed technical issues in the FNSPID project, like automating page navigation, filtering pop-ups, and recognizing new web elements. Using Selenium, we retrieved news headlines and full content from NASDAQ. This update extends coverage to January 19, 2025, compiling 216,308 high-quality news records. See Table~\ref{tab:stock_news} for details.

\begin{table*}[t]
    \centering
    \renewcommand{\arraystretch}{0.88} 
    \setlength{\tabcolsep}{10pt} 
    \begin{tabular}{|>{\centering\arraybackslash}m{2cm}|
                        >{\centering\arraybackslash}m{2cm}|
                        >{\centering\arraybackslash}m{2cm}|
                        >{\centering\arraybackslash}m{3.5cm}|
                        >{\centering\arraybackslash}m{3.5cm}|}
        \hline
        \textbf{Company} & \textbf{Total Days} & \textbf{Daily News Count} & \textbf{First Date} & \textbf{Last Date} \\
        \hline
        AAPL & 595 & 9365 & 2022/6/3 & 2025/1/18 \\
        \hline
        ABBV & 2669 & 6198 & 2013/1/3 & 2025/1/17 \\
        \hline
        AEP & 1224 & 1945 & 2009/10/9 & 2025/1/16 \\
        \hline
        AMT & 2196 & 6532 & 2009/12/2 & 2025/1/17 \\
        \hline
        AMZN & 307 & 5282 & 2023/3/9 & 2025/1/19 \\
        \hline
        APD & 1410 & 2213 & 2011/12/2 & 2025/1/17 \\
        \hline
        AVGO & 61 & 507 & 2023/12/16 & 2025/1/19 \\
        \hline
        BABA & 2573 & 9262 & 2014/11/7 & 2025/1/17 \\
        \hline
        BHP & 2354 & 4810 & 2009/10/21 & 2025/1/17 \\
        \hline
        BRK & 2087 & 9288 & 2009/10/19 & 2025/1/18 \\
        \hline
        CAT & 2366 & 5685 & 2009/11/4 & 2025/1/17 \\
        \hline
        CCI & 1156 & 1676 & 2010/2/23 & 2025/1/16 \\
        \hline
        CMCSA & 1877 & 4762 & 2017/4/6 & 2025/1/17 \\
        \hline
        COP & 2702 & 5711 & 2009/11/17 & 2025/1/17 \\
        \hline
        COST & 2883 & 7281 & 2010/2/17 & 2025/1/18 \\
        \hline
        CVX & 2217 & 9188 & 2016/8/3 & 2025/1/18 \\
        \hline
        D & 1691 & 2936 & 2010/4/20 & 2025/1/18 \\
        \hline
        DIS & 1667 & 9211 & 2019/6/13 & 2025/1/18 \\
        \hline
        DUK & 1925 & 3202 & 2010/4/26 & 2025/1/17 \\
        \hline
        GOOG & 1742 & 9226 & 2019/2/6 & 2025/1/19 \\
        \hline
        KO & 3143 & 8140 & 2009/10/14 & 2025/1/18 \\
        \hline
        LLY & 87 & 507 & 2023/12/16 & 2025/1/18 \\
        \hline
        LMT & 131 & 507 & 2023/12/15 & 2025/1/18 \\
        \hline
        MSFT & 631 & 9237 & 2022/4/26 & 2025/1/18 \\
        \hline
        NEE & 2383 & 4993 & 2010/7/24 & 2025/1/17 \\
        \hline
        NVDA & 828 & 9216 & 2021/8/17 & 2025/1/20 \\
        \hline
        PEP & 2519 & 5462 & 2009/11/10 & 2025/1/19 \\
        \hline
        PLD & 1434 & 2152 & 2010/6/24 & 2025/1/18 \\
        \hline
        RIO & 238 & 436 & 2012/10/31 & 2025/1/16 \\
        \hline
        SHW & 798 & 1218 & 2015/10/6 & 2025/1/15 \\
        \hline
        SO & 1577 & 2534 & 2010/6/22 & 2025/1/18 \\
        \hline
        TM & 2420 & 4083 & 2010/2/1 & 2025/1/18 \\
        \hline
        TSLA & 626 & 9212 & 2022/5/2 & 2025/1/19 \\
        \hline
        TSM & 2079 & 4398 & 2010/4/15 & 2025/1/19 \\
        \hline
        UNP & 1821 & 3084 & 2009/11/3 & 2025/1/18 \\
        \hline
        UPS & 2177 & 4217 & 2009/11/20 & 2025/1/18 \\
        \hline
        V & 2949 & 7445 & 2010/4/15 & 2025/1/19 \\
        \hline
        VALE & 1431 & 2015 & 2010/2/1 & 2025/1/17 \\
        \hline
        WMT & 1753 & 9154 & 2018/12/1 & 2025/1/19 \\
        \hline
        XOM & 1440 & 7253 & 2019/8/25 & 2025/1/18 \\
        \hline
        BAC & 96 & 500 & 2024/10/9 & 2025/1/18 \\
        \hline
        EQIX & 188 & 324 & 2015/7/21 & 2025/1/17 \\
        \hline
        HD & 146 & 500 & 2024/8/19 & 2025/1/17 \\
        \hline
        HON & 188 & 500 & 2024/5/14 & 2025/1/18 \\
        \hline
        JNJ & 127 & 500 & 2024/8/30 & 2025/1/18 \\
        \hline
        JPM & 89 & 500 & 2024/10/15 & 2025/1/17 \\
        \hline
        MA & 152 & 500 & 2024/8/1 & 2025/1/17 \\
        \hline
        META & 45 & 500 & 2024/12/6 & 2025/1/19 \\
        \hline
        PFE & 57 & 500 & 2024/10/18 & 2025/1/18 \\
        \hline
        PG & 182 & 500 & 2024/6/25 & 2025/1/18 \\
        \hline
        PSA & 146 & 241 & 2016/1/20 & 2025/1/17 \\
        \hline
        SHEL & 230 & 482 & 2024/2/1 & 2025/1/17 \\
        \hline
        TTE & 143 & 218 & 2024/2/13 & 2025/1/16 \\
        \hline
        UNH & 120 & 500 & 2024/9/5 & 2025/1/20 \\
        \hline
        VZ & 139 & 500 & 2024/8/21 & 2025/1/19 \\
        \hline
    \end{tabular}
    \caption{Stock News: Collection Periods and Total Days by Sector.}
    \label{tab:stock_news}
\end{table*}

\subsection{Multimodal Data Alignment}
To enhance stock prediction accuracy, we align time-stamped tweets, news articles, and stock prices as multimodal inputs. Integrating textual and numerical data improves predictive performance. For a comparison of daily tweet and news volumes, see Figure~\ref{fig:tweet_news_nums}.

\subsection{Text-Cleaning Pipeline}
Earlier sections detailed dataset attributes, including sampling frequency, time series count, and total observations. 
Here, we provide the text-cleaning pipeline, which ranks raw text and extracts summaries (see Algorithms~\ref{alg:news_processing} and \ref{alg:x_comment_ranking}) before converting them into embeddings (see Algorithm~\ref{alg:x_comment_embedding}). Tweets, being short, are not summarized. 

\begin{algorithm}[t]
\caption{News Ranking \& Summarization}
\label{alg:news_processing}
\begin{algorithmic}[1]
    \State \textbf{Input:} $\mathcal{D}_{\text{raw}}$ (JSONLs with $\{$Date, Text$\}$)
    \State \textbf{Output:} $\mathcal{D}_{\text{final}}$ (Summarized JSONLs)
    \State Load pretrained tokenizer and summarization model

    \For{each \texttt{file} in $\mathcal{D}_{\text{raw}}$}
        \State Extract $c$ (company code) from filename
        \State Read JSONL and group records by $t$ (date)
        \For{each $t$}
            \State Compute $\text{score}(x, c)$ for each $x$
            \State Select $x^* = \arg\max_x \text{score}(x, c)$
        \EndFor
        \State Store $\mathcal{D}_{\text{ranked}}$
    \EndFor

    \For{each \texttt{file} in $\mathcal{D}_{\text{ranked}}$}
        \For{each $x \in \mathcal{D}_{\text{ranked}}$}
            \State Truncate $x$ to $L_{\text{max}}=1024$
            \State Generate summary $\hat{x}$ using the pretrained model
            \State Store $\hat{x}$
        \EndFor
    \EndFor

    \State \textbf{Return} $\mathcal{D}_{\text{final}}$
\end{algorithmic}
\end{algorithm}

\begin{algorithm}[t]
\caption{X Comment Ranking}
\label{alg:x_comment_ranking}
\begin{algorithmic}[1]
    \State \textbf{Input:} $\mathcal{D}_{\text{raw}}$ (JSONLs with $\{$Date, Text$\}$)
    \State \textbf{Output:} $\mathcal{D}_{\text{ranked}}$ (Ranked JSONLs)
    
    \For{each \texttt{file} in $\mathcal{D}_{\text{raw}}$}
        \State Extract company code $c$ from filename
        \State Read JSONL, group records by date $t$
        \For{each $t$}
            \State Compute relevance score $\text{score}(x, c)$ for each $x$
            \State Select $x^* = \arg\max_x \text{score}(x, c)$
        \EndFor
        \State Store $\mathcal{D}_{\text{ranked}}$
    \EndFor

    \State \textbf{Return} $\mathcal{D}_{\text{ranked}}$
\end{algorithmic}
\end{algorithm}

\begin{algorithm}[t]
\caption{Text Embedding Generation}
\label{alg:x_comment_embedding}
\begin{algorithmic}[1]
    \State \textbf{Input:} $\mathcal{D}_{\text{ranked}}$ (JSONLs with $\{$Date, Text$\}$)
    \State \textbf{Output:} $\mathcal{D}_{\text{emb}}$ (Embedded JSONLs)
    
    \State Load pretrained tokenizer and encoder
    \State Set device and batch size

    \For{each \texttt{file} in $\mathcal{D}_{\text{ranked}}$}
        \State Read JSONL, extract and clean text
        \State Tokenize and batch non-empty entries
        \For{each batch}
            \State Compute embeddings via pretrained model
        \EndFor
        \State Replace text with embeddings, store $\mathcal{D}_{\text{emb}}$
    \EndFor

    \State \textbf{Return} $\mathcal{D}_{\text{emb}}$
\end{algorithmic}
\end{algorithm}

\subsection{Input-Output Flow in FTS-Text-MoE for Time Series Forecasting}
\begin{figure*}[t]
  \centering
  \includegraphics[width=1\linewidth]{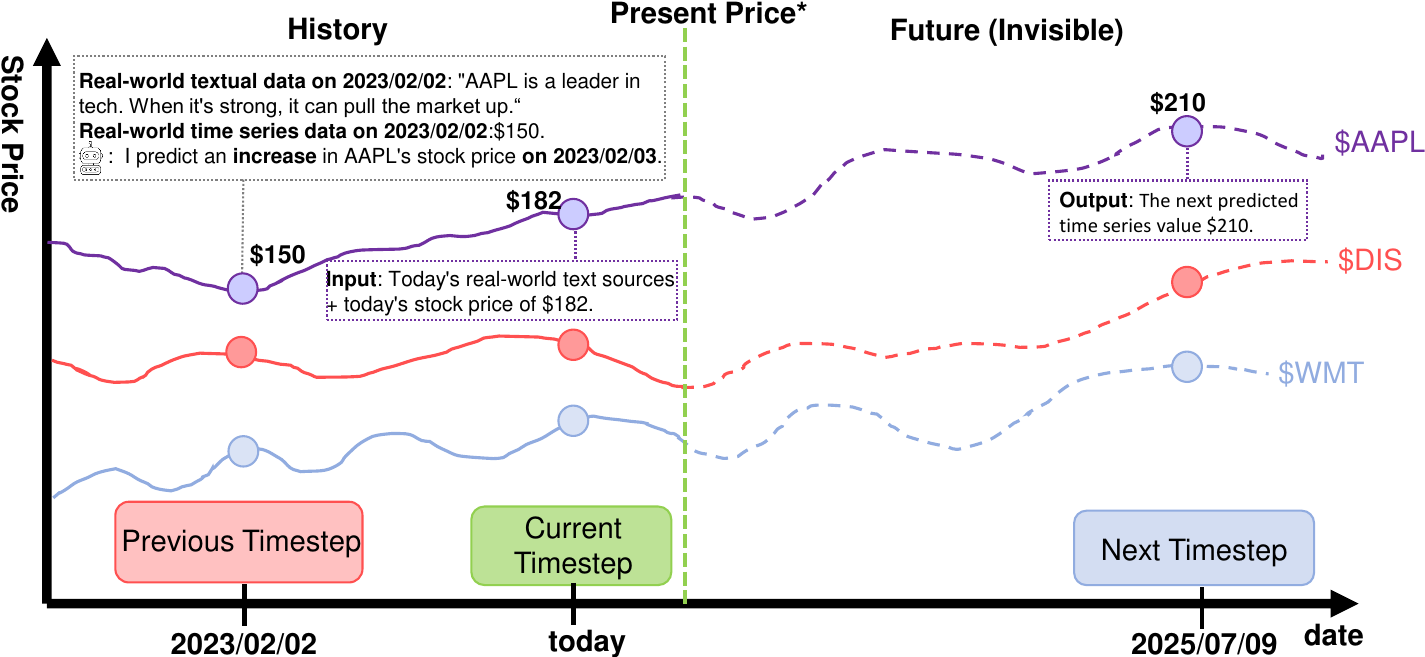}
   \vspace{-8mm}
  \caption{Illustration of Input and Output in Time Series Forecasting: FTS-Text-MoE takes the textual data and stock price of the current day as inputs and outputs the predicted stock price for the next day.}
  \label{fig:input_output}
\end{figure*}

Figure~\ref{fig:input_output} illustrates the input and output structure in time series forecasting using FTS-Text-MoE. 

\begin{figure*}[t]
  \centering
  \begin{minipage}[t]{0.30\textwidth}
    \centering
    \includegraphics[width=\textwidth]{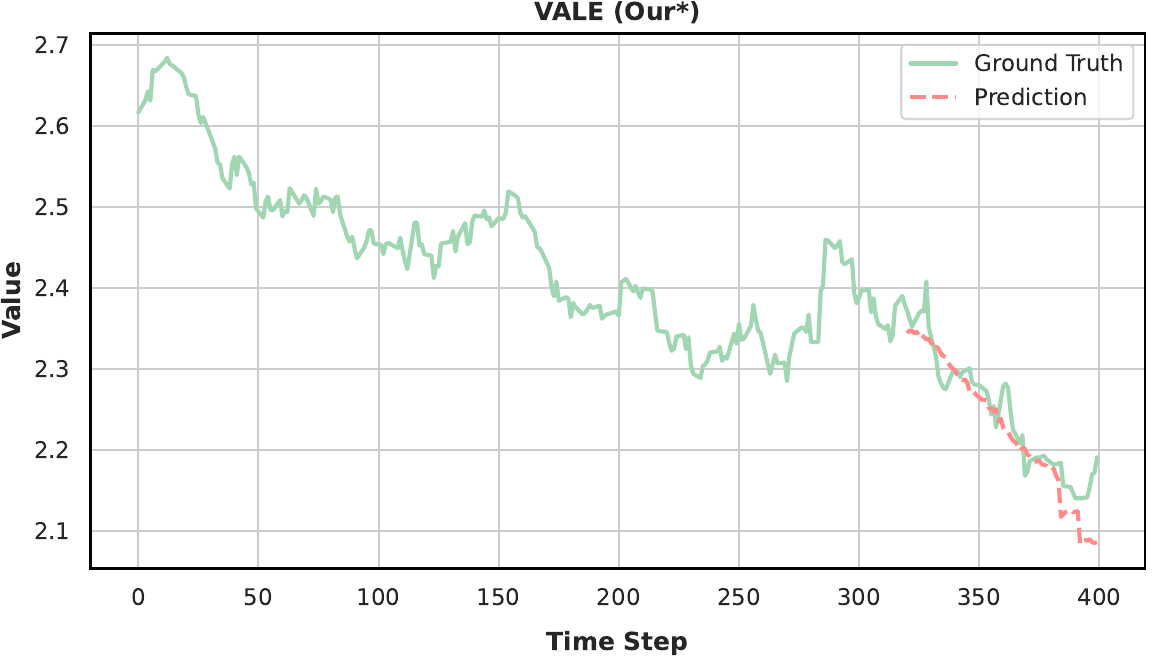}
    \label{fig:ts7}
  \end{minipage}
  \hfill
  \begin{minipage}[t]{0.32\textwidth}
    \centering
    \includegraphics[width=\textwidth]{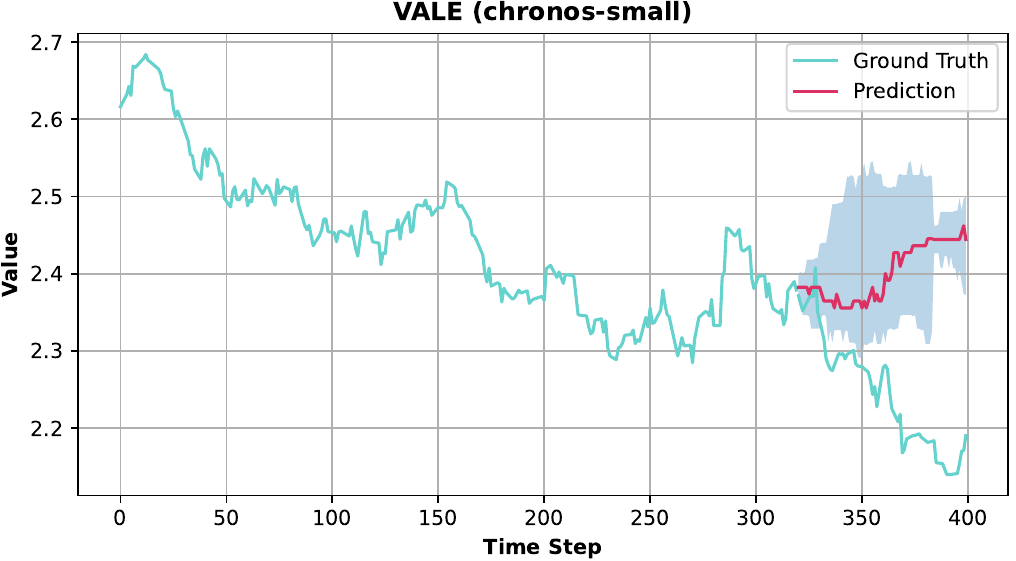}
    \label{fig:ts8}
  \end{minipage}
  \hfill
  \begin{minipage}[t]{0.32\textwidth}
    \centering
    \includegraphics[width=\textwidth]{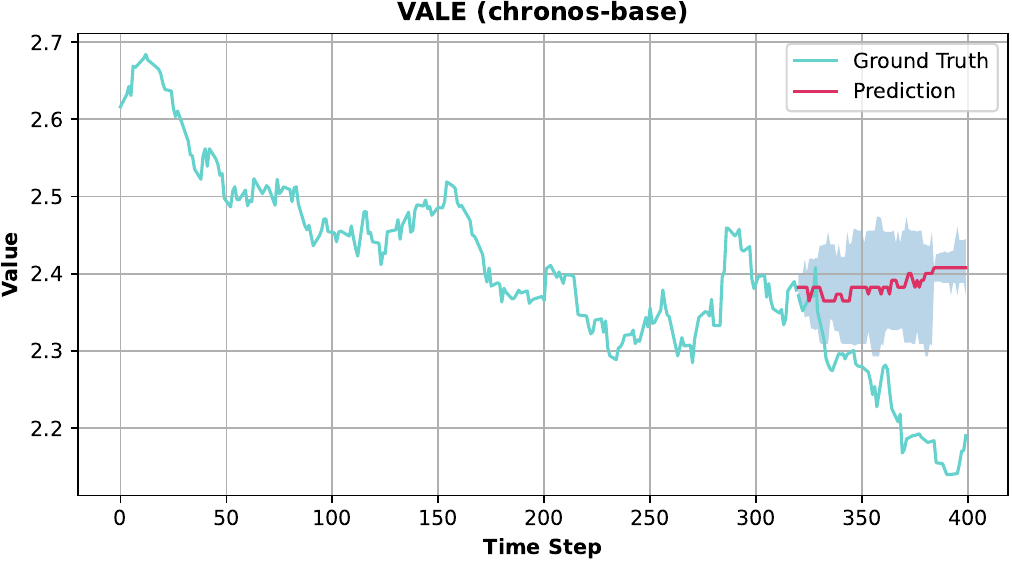}
    \label{fig:ts9}
  \end{minipage}
  
  \vspace{-5mm} 
  \begin{minipage}[t]{0.32\textwidth}
    \centering
    \includegraphics[width=\textwidth]{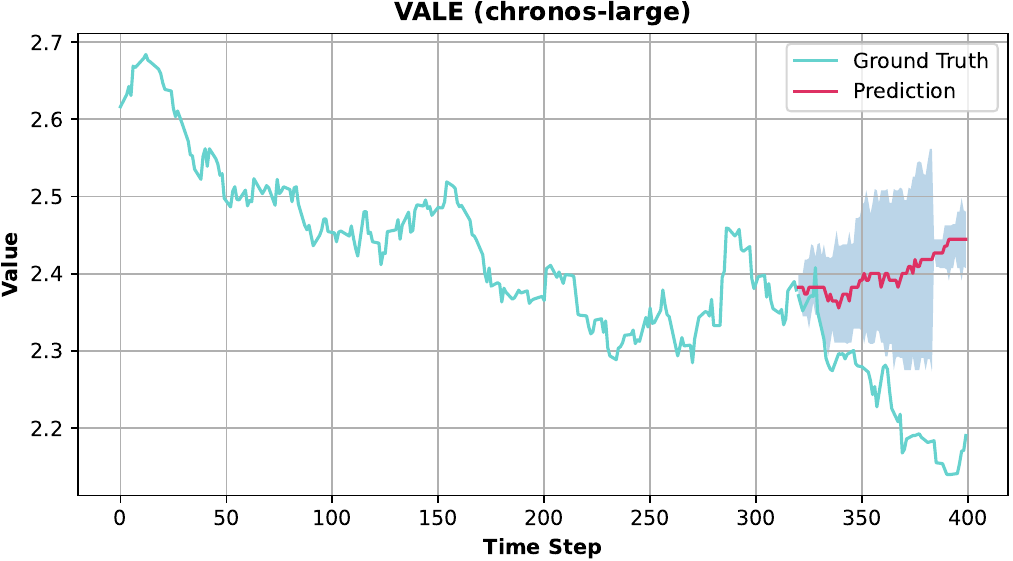}
    \label{fig:ts10}
  \end{minipage}
  \hfill
  \begin{minipage}[t]{0.32\textwidth}
    \centering
    \includegraphics[width=\textwidth]{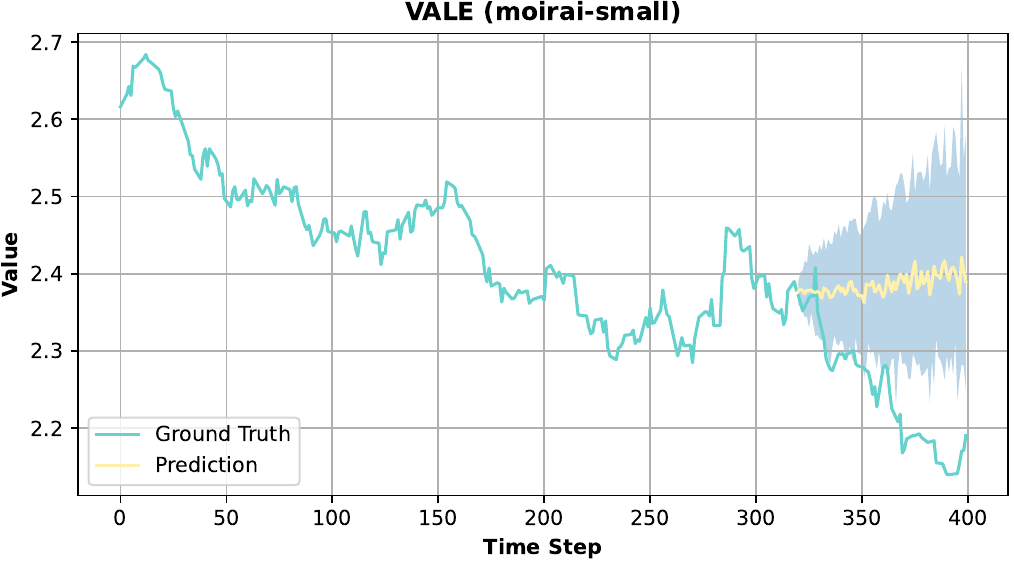}
    \label{fig:ts11}
  \end{minipage}
  \hfill
  \begin{minipage}[t]{0.32\textwidth}
    \centering
    \includegraphics[width=\textwidth]{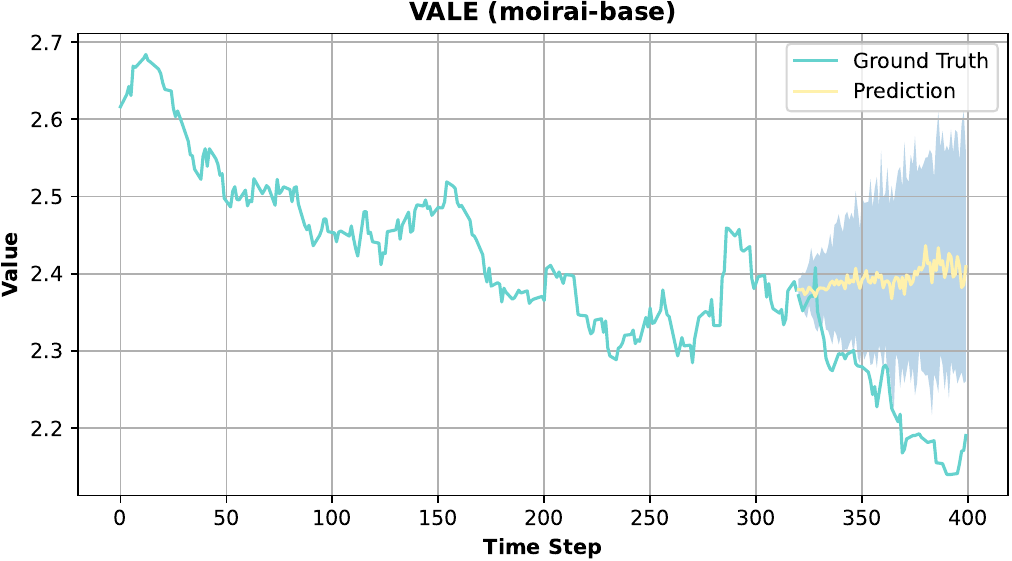}
    \label{fig:ts12}
  \end{minipage}
  \vspace{-8mm}
  \caption{Forecasting results of different models on the VALE stock time series with a forecast horizon of 80 and a context length of 320. The green/blue lines represent the ground truth, while the red/yellow lines indicate the model predictions.}
  \label{fig:VALE_6fig}
\end{figure*}

\subsection{Evaluation Metrics}
\label{Evaluation Metrics}
\paragraph{Error Metrics (Negative).} Model performance is evaluated using mean square error (MSE) and mean absolute error (MAE):

\small
\begin{align}
    \text{MSE} &= \frac{1}{H} \sum_{i=1}^{H} (x_{i} - \widehat{x}_{i})^2, \notag \\
    \text{MAE} &= \frac{1}{H} \sum_{i=1}^{H} |x_{i} - \widehat{x}_{i}|,
    \label{equ:metrics}
\end{align}
\normalsize

where $x_{i},\widehat{x}_{i} \in \mathbb{R}$ denote the ground truth and predicted values at time step $i$.

\paragraph{Overall (Positive).} Cumulative daily returns:

\small
\begin{equation}
    \text{Overall} = \sum_{i=1}^{n} R_{i}.
\label{Overall}
\end{equation}
\normalsize

\paragraph{Std. Dev. (Negative).} Standard deviation of returns, indicating risk:
\small
\begin{equation}
    \sigma = \sqrt{\frac{1}{n-1} \sum_{i=1}^{n} \left(R_{i}-\bar{R}\right)^{2}},
    \label{Std. Dev.}
\end{equation}
\normalsize
where the mean return is:
\small
\begin{equation}
    \bar{R} = \frac{1}{n} \sum_{i=1}^{n} R_{i}.
\end{equation}
\normalsize

\paragraph{Sharpe \cite{lo2002statistics} (Positive).} Risk-adjusted return:
\small
\begin{equation}
    \text{Sharpe} = \frac{\bar{R}_{p}-R_{f}}{\sigma_{p}} \times \sqrt{252}.
    \label{Sharpe}
\end{equation}
\normalsize
where $\bar{R}_{p}$ is the portfolio's average return, $R_{f}$ the risk-free rate, $\sigma_{p}$ the return standard deviation, and 252 the annualization factor.

These metrics provide a comprehensive evaluation of predictive accuracy and financial performance.

\section{Analysis of Financial Time-Series Results}

As shown in Figure \ref{fig:VALE_6fig}  (TTE stock time series data), Chronos and Moirai adopt a more conservative forecasting approach, with Moirai exhibiting the smoothest prediction curve. This suggests that Moirai prioritizes stability in high-uncertainty scenarios. In contrast, FTS-Text-MoE takes a more proactive approach, aiming to capture short-term fluctuations and better align with time series trends.


\end{document}